%% file: Main.tex
\documentclass{article}

    \PassOptionsToPackage{numbers, compress}{natbib}

\usepackage[preprint]{neurips_2024}

\usepackage[utf8]{inputenc} 
\usepackage[T1]{fontenc}    
\usepackage{hyperref}       
\usepackage{url}            
\usepackage{booktabs}       
\usepackage{amsfonts}       
\usepackage{nicefrac}       
\usepackage{microtype}      
\usepackage{xcolor}         

\usepackage{bibentry}
\usepackage{amsfonts}
\usepackage{subfigure}
\usepackage{multirow}
\usepackage{amsmath}

\usepackage{graphicx}
\usepackage{epsfig}
\usepackage{color}
\usepackage{epstopdf}
\usepackage{rotating}
\usepackage{caption}
\usepackage{float}
\usepackage{multicol}
\usepackage{amssymb}
\usepackage{graphicx}
\usepackage{bbding}
\usepackage[linesnumbered,ruled,vlined]{algorithm2e}
\usepackage{balance}
\usepackage{microtype}
\usepackage{graphicx}
\usepackage{subfigure}
\usepackage{booktabs}
\usepackage{multirow}
\usepackage{amsmath}
\usepackage{mathtools}
\usepackage{etoolbox}
\usepackage{cases}
\usepackage{enumitem}
\usepackage{xcolor}
\usepackage{subfigure}
\usepackage{cases}
\usepackage{dsfont}

\definecolor{lbcolor}{RGB}{13, 151, 175}

\newcommand{\method}{~RETURN~} 

\title{Retrieval-Augmented Purifier for Robust LLM-Empowered Recommendation}

\author{%
   Liangbo Ning \hspace{2em}  
  Wenqi Fan\thanks{Corresponding author: Wenqi Fan, Department of Computing, and 
 Department of Management and Marketing, The Hong Kong Polytechnic University.} \hspace{2em} 
  Qing Li 
  \\ 
  The Hong Kong Polytechnic University, Hong Kong SAR, China 
  \\ 
  \texttt{\{BigLemon1123,wenqifan03\}@gmail.com}, \texttt{qing-prof.li@polyu.edu.hk}
}

\begin{document}

\maketitle

\input{Section/Abstract}

\input{Section/Introduction}
\input{Section/Related_work}

\input{Section/Methodology}
\input{Section/Experiments}

\input{Section/Conclusion}

\balance
\bibliographystyle{ieeenat_fullname}
\bibliography{Reference}

\end{document}

%% file: Section/Abstract.tex
\begin{abstract}

Recently, Large Language Model (LLM)-empowered recommender systems have revolutionized personalized recommendation frameworks and attracted extensive attention. Despite the remarkable success, existing LLM-empowered RecSys have been demonstrated to be highly vulnerable to minor perturbations. To mitigate the negative impact of such vulnerabilities, one potential solution is to employ collaborative signals based on item-item co-occurrence to purify the malicious collaborative knowledge from the user's historical interactions inserted by attackers. On the other hand, due to the capabilities to expand insufficient internal knowledge of LLMs, Retrieval-Augmented Generation (RAG) techniques provide unprecedented opportunities to enhance the robustness of LLM-empowered recommender systems by introducing external collaborative knowledge. Therefore, in this paper, we propose a novel framework (\textbf{RETURN}) by retrieving external collaborative signals to purify the poisoned user profiles and enhance the robustness of LLM-empowered RecSys in a plug-and-play manner. Specifically, retrieval-augmented perturbation positioning is proposed to identify potential perturbations within the users' historical sequences by retrieving external knowledge from collaborative item graphs. After that, we further retrieve the collaborative knowledge to cleanse the perturbations by using either deletion or replacement strategies and introduce a robust ensemble recommendation strategy to generate final robust predictions. Extensive experiments on three real-world datasets demonstrate the effectiveness of the proposed RETURN.

\end{abstract}

%% file: Section/Introduction.tex
\section{Introduction}\label{sec:Introduction}

\begin{figure}[t]
\centering
\includegraphics[width=0.6\linewidth]{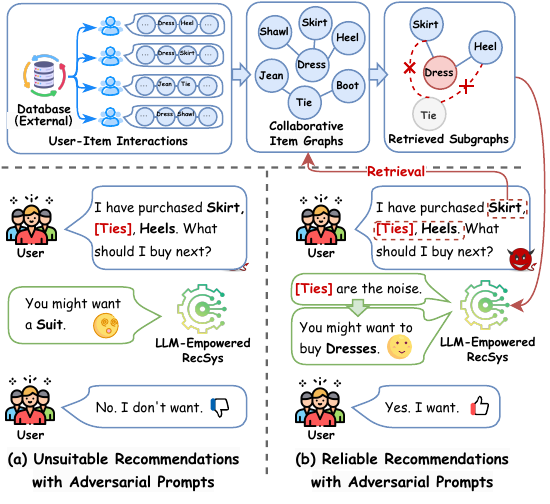}
\caption{
The illustration of the robust LLM-empowered RecSys by introducing an external database (i.e., collaborative item graph). 
The minor perturbations (e.g., item `\emph{Ties}') in the user's historical sequence (i.e., adversarial prompt) can mislead LLM-empowered recommender systems to understand the user's preference. 
With the help of the external data source, LLM-empowered recommender systems can identify the irrelevant item `\emph{Ties}' by retrieving relevant collaborative signals (i.e., retrieved subgraphs) from the collaborative item graphs, so as to purify the perturbations for the robust recommendation.
}
\label{fig:Illustration}
\end{figure}

In today's era of information explosion, recommender systems play a vital role in enhancing user experiences and influencing user decisions by filtering out irrelevant information in various applications such as streaming platforms (e.g., YouTube~\cite{covington2016deep,davidson2010youtube}, TikTok~\cite{liu2019user,cai2023two}) and e-commerce (e.g., Amazon~\cite{jin2024amazon}, Taobao~\cite{pfadler2020billion}). 
Technically, most existing representative recommendation methods aim to capture collaborative signals by modeling user-item interactions~\cite{fan2020graph,fan2019deep_dscf,he2020lightgcn}. 
Recently, large language models (LLMs) have been widely applied in real-life scenarios due to their powerful capabilities in language comprehension and generation, and rich store of open-world knowledge~\cite{qu2024survey,zhao2024recommender,ning2025surveywebagentsnextgenerationai}. 
For example, as one of the most famous AI chatbots in recent years, ChatGPT~\cite{achiam2023gpt} has showcased human-level intelligence with impressive logical reasoning, open-ended conversation, and personalized content recommendation abilities. 
To fully leverage the powerful capabilities of large language models, a significant amount of research has utilized LLMs to revolutionize recommender systems for next-generation RecSys~\cite{qu2024tokenrec,zhao2024recommender,lin2025can,wang2025automated}. 
For instance, 
\citet{geng2022recommendation} propose P5, which unifies various recommendation tasks by converting user-item interactions to natural language sequences, achieving outstanding recommendation performance due to the rich textual information that can help capture complex semantics for personalization and recommendations.

Despite the remarkable success, most existing LLM-empowered RecSys still encounter a key limitation, in which they have been demonstrated to be highly vulnerable to minor perturbations in the input prompt~\cite{ning2024cheatagent}, greatly constraining their practical applicability. 
Suppose that attackers might post products with enticing images and titles to attract user clicks on an e-commerce platform. 
Users are easily drawn to these clickbait products and interact with them, even though the content of these goods may not truly align with their preferences~\cite{wang2021clicks}. 
Such minor perturbations (e.g., irrelevant items) can easily lead the LLM-empowered RecSys to misunderstand the user preferences by capturing the collaborative knowledge from the user's historical interactions towards items. 
For example, 
as illustrated in Figure~\ref{fig:Illustration}, when perturbation item "ties" is inserted into the user's interaction sequence, the perturbed collaborative knowledge makes LLM-empowered RecSys struggle to discern whether the user is seeking men's clothing (i.e., "suits") or women's clothing (i.e., "dresses"), leading to inaccurate recommendation outcomes. That is due to the fact that attackers tend to add items that are irrelevant to users' behaviors for hindering collaborative knowledge learning~\cite{fan2021attacking,chen2022knowledge}. 
In order to defend such minor perturbations for robust recommendations, one of the promising solutions is to purify malicious collaborative knowledge from the user's historical interactions towards items in LLM-based recommender systems. 
In most recommender systems, collaborative graphs based on item-item co-occurrence are commonly employed as a collaborative signal to represent the relationships among items, 
where items frequently interacted together by different users are related (e.g., substitutable or complementary)~\cite{mcauley2015inferring,liang2016factorization}. 
Following the insertion of perturbations by attackers, item co-occurrence collaborative graphs as the external knowledge source can provide valuable evidence on whether such perturbations are relevant to other items in the user's interaction history and effectively filter out malicious collaborative signals (i.e., perturbed items), which can be achieved by retrieving subgraphs and examining the connection between the perturbation and the retrieved subgraphs.

Recently, to mitigate the problems usually caused by insufficient intrinsic knowledge of LLMs, 
including outdated knowledge, hallucination, and so on~\cite{rawte2023survey,gao2023retrieval,li2022survey,lyu2025crud}, retrieval-augmented generation (RAG) techniques~\cite{fan2024survey} have been proposed to expand the internal knowledge of large language models with an external database. 
The relevant knowledge is retrieved from the external database and employed to augment LLMs without changing the parameters of LLM backbone, achieving outstanding success for various knowledge-intensive domains such as open question answering~\cite{lewis2020retrieval}, medicine~\cite{ge2024development}, and finance~\cite{li2024alphafin,yepes2024financial}. 
For example, \citet{lewis2020retrieval} propose to utilize Wikipedia for knowledge retrieval and combine the retrieved documents with the input to augment the generation process, significantly improving the performance of LLMs for various complex tasks and mitigating the hallucination problem. 
In the context of RecSys, there exists a vast amount of publicly available external collaborative knowledge collected from various public platforms such as Amazon~\cite{linden2003amazon}, Yelp~\cite{luo2020finding}, and Steam~\cite{pathak2017generating}.
Given the success of expanding the internal knowledge of LLMs through the use of external databases to enhance their capabilities in a training-free manner, along with the abundant collaborative knowledge available in the RecSys community, RAG techniques provide unprecedented opportunities to enhance the robustness of LLM-empowered RecSys with external collaborative signals. 
For example, as shown in Figure~\ref{fig:Illustration}, LLM-empowered RecSys might generate an incorrect recommendation to a user who interacted with "skirt, ties, heels". 
To produce reliable recommendation results, a collaborative item graph (i.e., external databases) based on user-item interactions can be constructed to provide useful external collaborative knowledge for better understanding users' preferences in LLM-based recommender systems. 
LLM-based RecSys can purify the noisy users' online behaviors (i.e., perturbation "ties") by retrieving collaborative signals (i.e., subgraph) from the collaborative item graph for recommendation generation,  where items "skirt" and "heels"  rarely appear together with "ties" in most users' shopping behaviors.

To effectively take advantage of external collaborative signals from item-item collaborative graph, in this paper, a novel framework \textbf{RETURN} is proposed as a \underline{ret}rieval-a\underline{u}gmented pu\underline{r}ifier for e\underline{n}hancing the robustness of LLM-empowered recommender systems in a plug-and-play manner. 
Specifically, the users' historical sequences within the external databases are first encoded into collaborative item graphs to capture the extensive collaborative knowledge. 
After that, a retrieved-augmented perturbation positioning strategy is proposed to identify potential perturbations by retrieving relevant collaborative signals from the collaborative item graphs. 
Then, we further cleanse the potential perturbations within the user profile by using either deletion or replacement strategies based on the external collaborative item graphs. 
Finally, a robust ensemble recommendation strategy is proposed to guide the LLM-empowered RecSys to generate robust recommendation results. 
Our major contributions are summarised as follows: 
\begin{itemize}

    \item We introduce a novel strategy for denoising in LLM-empowered recommendation, in which training-free retrieval-augmented denoising strategy is proposed to leverage the collaborative signals of collaborative item graphs to purify the poisoned user profiles.

    \item     
    We propose a novel framework (\textbf{RETURN}) to enhance the robustness of LLM-empowered RecSys by harnessing collaborative signals from external databases in a plug-and-play manner. 
    Meanwhile, a robust ensemble recommendation is proposed to cleanse user profiles multiple times and generate robust recommendations by using a decision fusion strategy.

    \item We conduct extensive experiments on three real-world datasets to demonstrate the effectiveness of the proposed method. Comprehensive results indicate that RETURN can significantly mitigate the negative impact of the perturbations, highlighting the potential of introducing external collaborative knowledge to enhance the robustness of LLM-empowered recommender systems. 
    
\end{itemize}

The rest of this paper is organized as follows: 
Section~\ref{sec:Related_works} reviews multiple related studies. 
Section~\ref{sec:preliminary} provides the basic definition of the research problem, and the details of the proposed RETURN are presented in Section~\ref{sec:method}. Then, we conduct comprehensive experiments to investigate the effectiveness of RETURN in Section~\ref{sec:experiments}. 
Finally, we conclude the whole work in Section~\ref{sec:conclusion}. 

%% file: Section/Related_work.tex
\section{Related Works}\label{sec:Related_works}

\subsection{Defense Strategies for LLMs}
Numerous defense strategies have been devised to mitigate LLM vulnerabilities and safeguard against harmful information in LLM responses.
These methods are categorized into two main classes based on whether they are employed during training or inference. 

\noindent
\textbf{\emph{1) Defense in LLMs Training.}} 
The security of LLMs is significantly dependent on their training data, resulting in several defense strategies aimed at enhancing and purifying the training data~\cite{penedo2024refinedweb}. 
For example, 
\citet{wenzek2020ccnet} introduced CCNet, an automated pipeline designed to efficiently extract vast amounts of high-quality monolingual datasets from the Common Crawl corpus across various languages.
Beyond enhancing the quality of training data, adversarial training techniques~\cite{ijcai2021p591} are widely employed to guide LLMs towards appropriate behaviors by introducing adversarial perturbations into training examples to improve model robustness and performance~\cite{yao2024survey}.
For example, 
\citet{liu2020adversarial} introduced a general algorithm known as adversarial training for large neural language models (ALUM), aimed at enhancing the robustness of language models. ALUM enhances model resilience by regularizing the training objective via incorporating perturbations within the embedding space and focusing on maximizing adversarial loss.
\citet{wang2019improving}) propose a simple yet effective adversarial training method that incorporates adversarial perturbations into the output embedding layer during model training. 
\citet{li2021token} employs a token-level accumulated perturbation vocabulary to initialize the adversarial perturbations and use a token-level normalization ball to regulate the generated perturbations for virtual adversarial training~\citep{chen2020seqvat}.

\noindent
\textbf{\emph{2) Defense in LLMs Inference.}}
The large scale of parameters of LLMs renders their retraining or fine-tuning processes both time-consuming and computationally expensive. Therefore, training-free defense methods during inference have drawn considerable attention~\cite{yao2024survey}. 
For example, 
\citet{kirchenbauer2023reliability} and \citet{jain2023baseline} undertake extensive experiments to evaluate the effectiveness of different defense methods, such as perplexity-based detection, retokenization, and paraphrasing. 
\citet{li2023text} introduce an adversarial purification method that masks input texts and leverages masked language models~\cite{kenton2019bert} for text reconstruction.
\citet{wei2023jailbreak} and \citet{mo2023test} propose enhancing model robustness through contextual demonstrations. 
\citet{wang2023rmlm} propose RMLM, aimed at countering attacks by confusing attackers and correcting adversarial contexts stemming from malicious perturbations.
\citet{helbling2023llm} incorporate generated content into a predefined prompt and utilize another LLM to analyze the text and assess its potential harm.

\subsection{LLM-Empowered Recommender Systems}\label{appendix:related_work}
Currently, LLMs are widely employed in enhancing the capabilities of recommender systems due to their powerful language understanding, logical reasoning, and generation abilities. 
These studies can be generally divided into three categories based on the item information utilized. 

\noindent
\textbf{\emph{1) ID-Based LLM-Empowered Recommender Systems.}}
ID-based LLM-empowered recommender systems represent an item with a numerical index and use the item IDs for recommendations~\cite{geng2022recommendation,zheng2024adapting}. 
For example, 
\citet{geng2022recommendation} propose P5, which unifies various recommendation tasks by converting user-item interactions to natural language sequences. P5 introduces whole-word embedding to represent the token IDs, bridging the gap between large language models and recommender systems. 
\citet{zheng2024adapting} propose a learning-based vector quantization method for assigning meaningful item indices for items and introduce specialized tasks to facilitate the integration of collaborative semantics in LLMs, leading to an effective adaptation to recommender systems.

\noindent
\textbf{\emph{2) Text-Based LLM-Empowered Recommender Systems.}}
To effectively harness the natural language understanding and generation capabilities of LLMs, text-based LLM-empowered recommender systems primarily leverage textual information such as item titles and item descriptions for recommendation~\cite{du2024enhancing,bao2023tallrec}. 
For example, 
\citet{bao2023tallrec} introduce TALLRec, a novel tuning paradigm designed to tailor LLMs for recommendation tasks effectively, guides the model to assess user interest in a target item by analyzing their historical interactions that encompass textual descriptions like item titles. 
\citet{du2024enhancing} propose a novel LLM-based approach for job recommendation that enhances user profiling for resume completion by extracting both explicit and implicit user characteristics based on users' self-description and behaviors. A GANs-based method is introduced to refine the representations of low-quality resumes, and a multi-objective learning framework is utilized for job recommendations.

\noindent
\textbf{\emph{3) Hybrid LLM-Empowered Recommender Systems.}}
These approaches effectively integrate both textual information and ID-based knowledge to generate recommendations~\cite{liao2024llara,ren2024enhancing}. 
For example, 
\citet{ren2024enhancing} leverage text-format knowledge from LLMs and item IDs to enhance recommendation performance, along with a novel alignment training method and an asynchronous technique to refine LLMs' generation process for improved knowledge augmentation and accelerated training. 
\citet{liao2024llara} propose a novel hybrid prompting approach that integrates ID-based item embedding generated by traditional RecSys with textual item features. Besides, LLaRA utilizes a projector to align traditional recommender ID embeddings with LLM input space and incorporates a curriculum learning strategy to gradually train the model to integrate behavioral knowledge from traditional sequential recommenders, thereby enhancing recommendation performance seamlessly.

\subsection{Denoising for Traditional Recommender Systems}
With the development of RecSys, a growing body of research has focused on their vulnerability to noisy data, subsequently driving the advancement of various denoising approaches to improve system robustness~\cite{wang2021denoising,lin2023self,zhang2020gcn,tian2022learning}. 
For example, 
GraphRfi~\cite{zhang2020gcn} proposes an innovative end-to-end framework that integrates Graph Convolutional Networks (GCN) and neural random forests to simultaneously enhance robust recommendation accuracy and fraudster detection. By leveraging user reliability features and prediction errors of RecSys, GraphRfi effectively mitigates the impact of shilling attacks~\cite{si2020shilling,gunes2014shilling}. 
LoRec~\cite{zhang2024lorec} proposes to enhance the robustness of sequential recommender systems against poisoning attacks by integrating the open-world knowledge of large language models. 
Through LLM-Enhanced Calibration, LoRec employs a user-wise reweighting strategy to generalize defense mechanisms beyond specific known attacks, effectively mitigating the impact of fraudsters. 
LLM4DASR~\cite{wang2024llm4dsr} introduces an LLM-assisted denoising framework for sequential recommendations, combining self-supervised fine-tuning with uncertainty estimation to address output quality challenges. 
This model-agnostic framework effectively identifies and corrects noisy interactions, enhancing recommendation performance across various models.

\subsection{Difference between Existing Denoising Approaches and RETURN}
Despite the presence of existing denoising techniques, they are fundamentally different from our approach in terms of task formulation and technical details: 

    \noindent 1) \textbf{Denoising for different phases. }
    Existing denoising methods primarily focus on purifying the training set and ensuring accurate representation learning for RecSys to mitigate the impact of shilling attacks during training, assuming that user historical interactions during inference contain no perturbations. 
    However, during the inference phase, users may still be attracted to clickbait items and interact with them, leading to perturbations that do not align with their true preferences. 
    Moreover, studies~\cite{ning2024cheatagent} have highlighted the vulnerability of LLM-empowered recommender systems during inference, where even a \textbf{well-trained} LLM-based RecSys frequently produces inaccurate recommendations for users affected by poisoned interactions.
    In other words, even after the training set has been purified, if a user inadvertently interacts with a few clickbait or disliked items during inference, LLM-empowered RecSys may still misinterpret the user's preferences and generate unsatisfied recommendations. 
    In this paper, we assume that the LLM-empowered RecSys is well-trained, while user interaction sequences may contain noise or adversarial perturbations during inference. 
    In other words, RETURN is designed to address the \textbf{inference-phase vulnerability of LLM-empowered RecSys} and enhance their robustness, which is fundamentally different from the objective of previous denoising methods. 
    Additionally, RETURN can be seamlessly integrated with prior denoising approaches. For instance, existing methods can be employed to cleanse the training set and train a powerful LLM-empowered RecSys, while RETURN ensures the robustness of the RecSys during the inference phase. 
    
    \noindent 2) \textbf{Novel purification techniques based on external collaborative signals. } 
    Existing denoising methods primarily rely on leveraging the characteristics of perturbations~\cite{wang2021denoising,zhang2020gcn,tian2022learning} or the open-world knowledge of LLMs~\cite{zhang2024lorec,wang2024llm4dsr} to identify perturbations within the training set, largely overlooking the potential of external collaborative knowledge. 
    With the advancement of recommender systems, numerous publicly available datasets have been introduced to evaluate algorithm performance. These datasets offer abundant external collaborative signals that can be leveraged to purify perturbations within user historical interactions. 
    Specifically, after attackers introduce perturbations, collaborative graphs based on item-item co-occurrence can be constructed from the external database and leveraged to assess whether these perturbations are consistent with other items in the user’s interaction history, thereby effectively filtering out malicious collaborative signals (i.e., perturbed items). 
    Therefore, RETURN effectively extracts collaborative knowledge from external databases to purify the user historical interactions in a plug-and-play manner, providing a promising solution for enhancing the robustness of LLM-empowered RecSys.

%% file: Section/Methodology.tex
\section{Problem Statement}\label{sec:preliminary}
\subsection{Notation and Definition}
The objective of recommender systems is to capture the users' preferences from their historical interactions, such as browsing, clicking, and purchasing. 
In the era of LLMs,  the recommendation task is usually converted to the natural language format, consisting of user $u_i \in U=\{u_1,u_2,...,u_{|U|}\}$, and user's interaction history (also called user's profile) $\mathcal{I}_{u_i}=[I_1,I_2,...,I_{|\mathcal{I}_{u_i}|}]$, and a recommendation prompt $\mathcal{P}=[p_1,p_2,...,p_{|\mathcal{P}|}]$, where 
$p_i$ is the textual token used to guide the RecSys $\mathcal{R}_{\theta}$ to generate recommendations.  
$I_i \in \mathcal{I}=\{I_1,I_2,...,I_{|\mathcal{I}|}\}$ is the interacted item from the item pool $\mathcal{I}$ of user $u_i$. 
Based on the above definition, a textual recommendation query can be represented as $\textbf{x}=[\mathcal{P} \circ u_i \circ \mathcal{I}_{u_i}]$, where $\circ$ represents inserting the information of user $u_i$ and the corresponding interaction list $\mathcal{I}_{u_i}$ into the designated position of prompt $\mathcal{P}$. 
For example, as shown in Figure~\ref{fig:framework}, after inserting the user information and item interaction sequence into the prompt $\mathcal{P}$, the specific input used for recommendation can be denoted by: 
\begin{equation}
    \begin{aligned}
         \mathcal{P}=&[ \text{what, is, the, top, recommended, item, for, } [\underline{User\_235}], \\ 
        & ~~~ \text{who, interacted, with,} [\underline{item\_123,...,item\_928}], ? ], 
    \end{aligned}
\end{equation}

\noindent where $u_i=[User\_235]$ and $\mathcal{I}_{u_i}=[item\_123,...,item\_928]$ are the specific user and the historical interactions of user $u_i$, respectively. 
In different LLM-empowered RecSys, $\mathcal{I}_{u_i}$ can take various forms, such as numeric IDs~\cite{geng2022recommendation} or item titles~\cite{bao2023tallrec} for recommendations. 
Assume the target item is $\textbf{y}$, the performance of the LLM-empowered RecSys can be defined by: 
\begin{align}
\mathcal{D} (\mathcal{R}_{\theta}(\textbf{x}), \textbf{y}), 
\end{align}

\noindent where $\mathcal{D}$ evaluates the discrepancy between the generated recommendations $\mathcal{R}_{\theta}(\textbf{x})$ and the ground truth. 
During training, the negative log-likelihood function could be used as the $\mathcal{D}$, while during inference, the Hit Ratio or Normalized Discounted Cumulative Gain (NDCG)~\citep{geng2022recommendation} could be employed to evaluate the recommendation performance.

\subsection{Vulnerabilities of LLM-Based RecSys}
The vulnerabilities of LLM-based RecSys refer to the phenomenon where the model's recommendation outcomes vary significantly due to minor perturbations in the input~\cite{ning2024cheatagent}. 
Such vulnerabilities significantly deteriorate the overall user experience and compromise the effectiveness of RecSys. 
For example, assuming a user $u_i$ inadvertently clicks on some items they are not actually interested in, leading to a change in their interaction history from $\mathcal{I}_{u_i}$ to $\mathcal{\hat I}_{u_i}=\mathbb{I}(\mathcal{I}_{u_i} \circ \delta|s)$, where $\mathbb{I}(\mathcal{I}_{u_i} \circ \delta|s)$ represent to insert perturbation $\delta$ into user's profile $\mathcal{I}_{u_i}$ at position $s$. The vulnerability of LLM-empowered RecSys may lead the system to recommend items that the user is not interested in, consequently resulting in a decline in user experience. 
As an attacker, the perturbations $\delta$ can be generated intentionally by optimizing the following equation: 
\begin{align}
\delta = \underset{\delta: |\delta| \le  {\triangle}}{\arg \max} ~\mathcal{D}(\mathcal{R}_{\theta}(\hat{\textbf{x}}), \textbf{y}),
\end{align}

\noindent where $\hat{\textbf{x}}=[\mathcal{P} \circ u_i \circ \mathcal{\hat I}_{u_i}]$ is the perturbed input and $\triangle$ constrains the magnitude the perturbations. 
$\mathcal{D}$ evaluates the discrepancy between the generated recommendations $\mathcal{R}_{\theta}(\textbf{x})$ and the ground truth. 

\subsection{Robust LLM-based Recommendation}
The primary objective of robust LLM-based recommendations is to prevent the negative impact of the perturbations contained in the users' profiles, thereby enhancing the system's reliability and robustness. 
There are mainly two approaches to achieve this goal: adversarial training-based methods~\cite{xhonneux2024efficient} and training-free methods~\cite{varshney-etal-2024-art}. 
Adversarial training-based methods intentionally create multiple perturbed training samples to guide the RecSys in learning patterns of perturbations, thereby improving system robustness. 
However, these methods usually retrain or fine-tune the whole RecSys, which is extremely time-consuming due to the large number of trainable parameters in LLMs. 
Consequently, this paper primarily concentrates on the training-free methods, which improve the model's robustness without introducing additional changes in model parameters. 
Specifically, when the users' profiles contain adversarial perturbations, we aim to accurately identify and filter out these perturbations to ensure the appropriate recommendations for users during the inference process. 
Mathematically, if the input with adversarial perturbations is denoted by $\hat{\textbf{x}}$, we aim to cleanse the input for robust recommendations, formulated as follows: 
\begin{align}
\bar{\textbf{x}} = {\arg \min} ~\mathcal{D}(\mathcal{R}_{\theta}({\mathbb{C}({\hat{\textbf{x}}})}), \textbf{y}),
\end{align}

\noindent
where $\mathbb{C}(\hat{\textbf{x}})$ represents purifying input $\hat{\textbf{x}}$ containing perturbations into a benign prompt $\bar{\textbf{x}}$.

\section{Methodology}\label{sec:method}
\begin{figure*}[t]
\centering
\includegraphics[width=1\linewidth]{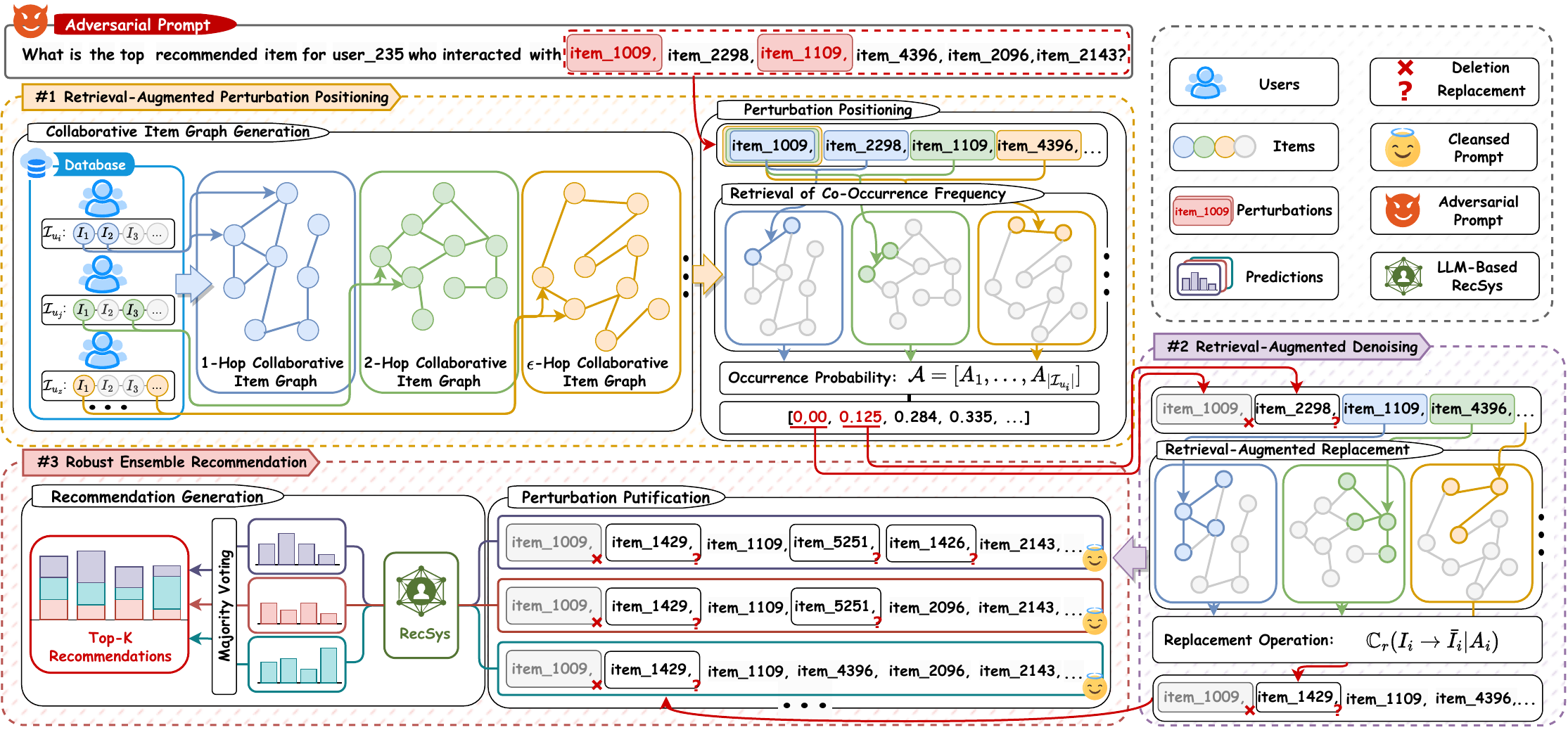}
\vskip -0.1in
\caption{
The overall framework of the proposed RETURN. The user interaction sequences in the external database are first converted to multi-hop collaborative item graphs. 
The occurrence probability of each item is computed based on the collaborative item graph for perturbation positioning. Finally, we purify the input prompt by retrieving collaborative signals from the collaborative item graphs for robust recommendation generation. 
}
\label{fig:framework}
\end{figure*}

\subsection{An Overview of the RETURN}
RETURN is proposed to leverage the collaborative knowledge of users within external databases to filter out adversarial perturbations, thereby enhancing the robustness of the existing LLM-based RecSys. 
As shown in Figure~\ref{fig:framework}, RETURN mainly contains three components: 
Retrieval-Augmented Perturbation Positioning, Retrieval-Augmented Denoising, and Robust Ensemble Recommendation. 
First, we convert the user interaction sequences within the external database to collaborative item graphs to encode the collaborative knowledge without introducing additional training processes. 
After that, the probability of each item in the user profile being a perturbation is computed by retrieving collaborative signals from collaborative item graphs. 
Second, retrieval-augmented denoising filters out the potential perturbations in the user profiles using either deletion or replacement strategies based on the collaborative signals of the generated item graphs. 
Finally, robust ensemble recommendation purifies input query multiple times and adopts an ensemble strategy to generate the final recommendations.

\subsection{Retrieval-Augmented Perturbation Positioning}

To mitigate the negative impact of perturbations, the first crucial step is to accurately locate the perturbations from extensive interactions within user profiles. 
To achieve this goal, we propose to use collaborative item graphs to encode the collaborative signals from users in the external database and retrieve relevant collaborative knowledge for perturbation positioning. 
By encoding the user's interaction history into collaborative item graphs, we can clearly understand the relationships between items~\cite{zhang2024denoising}, and such strong collaborative signals can provide evidence for subsequent denoising processes. 
Furthermore, this approach enables the direct use of a one-hot vector for information retrieval, eliminating the need to explicitly train a universal retriever as required by other RAG techniques, thereby improving efficiency. 

\subsubsection{Collaborative Item Graph Generation}

Let $\mathcal{E}=\{U_{\mathcal{E}}, \mathcal{I}_{\mathcal{E}}\}$ be the external database, where $U_{\mathcal{E}}=\{u_1,u_2,...,u_{\mathcal{E}}\}$ and $\mathcal{I}_{\mathcal{E}}=\{\mathcal{I}_{u_1},\mathcal{I}_{u_2},...,\mathcal{I}_{u_{\mathcal{E}}}\}$ denote the users and their interaction sequences, respectively. 
The most straightforward method of generating collaborative item graphs is to count the occurrence frequency of $i$-th and $j$-th items appearing together in the historical interaction of the same user. 
However, such a vanilla strategy overlooks the temporal relationships among items, which are significantly crucial for subsequent denoising processes. 
For instance, mobile phones and phone cases are usually interacted with consecutively by users, whereas mobile phones and furniture are typically not sequentially interacted with by users. 
During the denoising process, if a user consecutively interacts with both mobile phones and furniture, there is a high likelihood of perturbations in the user's historical interactions. 
Therefore, to provide precise collaborative signals, we also consider the gap between two items and generate a set of multi-hop collaborative item graphs, which encode not only the relevance between items but also the temporal relationships of items.
Given the external database $\mathcal{E}=\{U_{\mathcal{E}}, \mathcal{I}_{\mathcal{E}}\}$, the multi-hop collaborative item graph can be represented as:
\begin{align}
    \mathcal{G}_{\epsilon} =\{\mathcal{I},\mathcal{C}^{\epsilon}_{I_i,I_j}\}
    =\{\mathcal{I}, \mathbb{T}(\mathcal{C}^{\epsilon}_{I_i,I_j}|u_z, \epsilon) \}, 
    \label{eq:generate_graph}
\end{align}
\noindent
where the $\epsilon$-hop collaborative item graph contains nodes $\mathcal{I}$ and edges $\mathcal{C}^{\epsilon}_{I_i,I_j}$, respectively. 
The edge $\mathcal{C}^{\epsilon}_{I_i,I_j}$ stores the co-occurrence frequency of two items.  $\mathbb{T}(\cdot)$ is a counting function. 
If two items $I_i$ and $I_j$ appear simultaneously within the historical interactions of user $u_z$ with a gap of $\epsilon$ items between them, the co-occurrence frequency $\mathcal{C}^{\epsilon}_{I_i,I_j}$ is increased by one, defined by: 
\begin{align}
    \mathbb{T}(\mathcal{C}^{\epsilon}_{I_i,I_j}|u_z, \epsilon) \} =  
    \begin{cases}
    \mathcal{C}^{\epsilon}_{I_i,I_j} + 1, & \text{ if } I_i, I_j \in \mathcal{I}_{u_z}, |i-j|=\epsilon,  \\
    \mathcal{C}^{\epsilon}_{I_i,I_j},  & \text{ otherwise }.
    \end{cases}
\end{align}

\subsubsection{Perturbation Positioning} 

After encoding the external users' collaborative knowledge into collaborative item graphs, the next step is to locate the perturbations within the input query based on the generated graphs. 
Specifically, if one item has never appeared together with the remaining items in the user's historical interactions based on the collaborative item graphs derived from the majority of users' behavior, this indicates that such item is unrelated to the other items the user has interacted with. 
Thus, the likelihood of this item appearing within the user's interaction history is minimal, and the occurrence of such a low-probability event strongly implies that this item is likely introduced as a perturbation by an attacker.
Therefore, to accurately locate the potential perturbations, 
we propose to retrieve co-occurrence frequency from the collaborative item graphs and assess the probability of each item appearing within the user's interaction history. 

Given the multi-hop collaborative item graphs $\mathcal{G}_{\epsilon}$ and the user's historical interactions $\mathcal{I}_{u_i}=[I_1,I_2,...,I_{|\mathcal{I}_{u_i}|}]$, the co-occurrence frequency for a pair of items can be defined as:  
\begin{align}
o_{i,j} = {\frac{\mathbb{R}(I_i,I_j|\mathcal{G}_{\epsilon})}{\sum_{z \in [1,|\mathcal{I}|]}^{} {\mathbb{R}(I_i,I_z|\mathcal{G}_{\epsilon})}}},
\label{eq:occurrence_score}
\end{align}

\noindent
where $\epsilon=|j-i|$ is the gap between $i$-th and $j$-th items and $\mathbb{R}(I_i,I_j|\mathcal{G}_{\epsilon})$ represent to retrieve the co-occurrence frequency between $i$-th and $j$-th items from the $\epsilon$-hop collaborative item graph. 
By traversing each pair of items in the user's historical interactions, the occurrence probability of each item is denoted by:
\begin{equation}
    \begin{aligned}
        \mathcal{A}=[A_1,...,A_{|\mathcal{I}_{u_i}|}], \\
        A_i = {\textstyle \sum_{j=1, j \ne i}^{|\mathcal{I}_{u_z}|}} {o_{i,j}}. 
    \end{aligned}
    \label{eq:occurrence_probability}
\end{equation}
\noindent 
A smaller value of $A_i$ indicates a lower probability of the current item co-occurring with other items, making it more likely to be a perturbation.

\subsection{Retrieval-Augmented Denoising}
Once the occurrence probability of each item has been computed, it is necessary to purify the input query based on such collaborative knowledge. 
However, directly removing numerous items that may be perturbations usually leads to the RecSys failing to capture the user's preferences accurately since there are limited remaining interactions.  
To mitigate the negative impact of perturbations while maintaining the integrity of user interaction sequences, a hybrid strategy is proposed to eliminate a small subset of items that are most likely perturbations and replace the remaining potential perturbations with items that align with the user's preferences. 

If an item's occurrence probability $A_i=0$, it indicates that this item has never co-occurred with the other items in the user's historical interactions based on the collaborative signals of most users from external databases. Thus, this item is highly likely a perturbation inserted by attackers due to its lack of relevance to the user's other interaction items, and its deletion typically helps RecSys accurately capture the user's genuine preferences. 
Mathematically, given the interaction history $\mathcal{I}_{u_i}$ of the user $u_i$ and the occurrence probability $\mathcal{A}$, we first delete the most likely perturbation items whose occurrence probability is zero, defined by: 
\begin{align}
    \mathbb{C}_d({I}_i \to \emptyset|{A}_i) = \mathds{1}(I_i, A_i), 
\end{align}
\noindent
where $\mathds{1}(I_i, A_i)$ represents to execute deletion operation when $A_i=0$ and preservation otherwise. 

After removing items with $A_i=0$ that are most likely perturbations, some items usually remain in the user's historical interactions with very low but non-zero occurrence probabilities. 
Simply deleting these items usually results in sparse user-item interactions, and such limited collaborative knowledge leads to cold start issues~\cite{wei2021contrastive}, hindering RecSys from capturing user preferences effectively. 
To maintain the integrity of user interaction history, a retrieval-augmented replacement strategy is proposed to replace the remaining potential perturbations with items that align with user preferences. 
Specifically, all items that have co-occurred with the other remaining items in the user's interaction history are retrieved from the collaborative item graphs, and the item that shows the highest co-occurrence frequency is considered the prime candidates that best align with the current user preferences among all the retrieved items.
Given the user's historical interactions $\mathcal{I}_{u_i}=[I_1,...,I_{\mathcal{I}_{|u_i|}}]$ and the potential perturbation $I_i$ with the low occurrence probability, the replacement operation is defined by: 
\begin{align}
    \mathbb{C}_r(I_i \to {\bar I}_i|A_i) = \underset{{\bar I}_i}{\arg\max} \sum_{j=1, j \ne i}^{|\mathcal{I}_{u_i}|} \frac{ \mathbb{R}(I_j|\mathcal{G}_{\epsilon}) \cdot A_j } {\sum_{z \in [1,|\mathcal{I}|]} \mathbb{R}(I_j,I_z|\mathcal{G}_{\epsilon})} ,
\label{eq:replacement}
\end{align}
\noindent
where $\mathbb{C}_r(I_i \to {\bar I}_i|A_i)$ represents to replace the potentially perturbed item $I_i$ with alternative items ${\bar I}_i$ that better align with user preferences. 
$\mathbb{R}(I_j|\mathcal{G}_{\epsilon})$ represent to retrieve the co-occurrence frequency between item $I_j$ and all items that have co-occurred with $I_j$ from the item pool $\mathcal{I}$ based on the $\epsilon$-hop collaborative item graph $\mathcal{G}_{\epsilon}$, where $\epsilon=|j-i|$. 
$A_j$ is considered as a weight, where a larger $A_j$ indicates a closer alignment between the current item $I_j$ and the user's preference, thus resulting in greater weights assigned to items that are likely to co-occur with it. 

\subsection{Robust Ensemble Recommendation}
Due to the uncertainty regarding the number of perturbations, determining the extent of purification applied to the user's historical interactions is a challenging task. 
Excessive modification of items leads to difficulties in capturing the user's intrinsic preferences, thereby diminishing the recommendation performance. 
Conversely, the limited purification of items results in perturbations still existing in user profiles, making it challenging to enhance the robustness of RecSys. 
To tackle this challenge, we propose a robust ensemble recommendation approach. 
Specifically, we first randomly purify varying numbers of items in the user's profile and generate a set of cleansed inputs. 
These cleansed prompts are fed into the LLM-based RecSys, and the final recommendations are obtained by adopting a voting mechanism~\cite{lam1997application}. 
Technically, we randomly sample an integer $n$ from a normal distribution. 
Top-$n$ items $\mathcal{\hat I}_{u_i}^n=[I^1,I^2,...,I^n]$ with the lowest occurrence probabilities are identified from the user's historical interactions based on $\mathcal{A}$ and deletion $\mathbb{C}_d$ or substitution $\mathbb{C}_r$ operations are performed on these items. 
The purified user profile is defined by: 
\begin{align}
    \mathcal{\bar I}_{u_i} =[\mathbb{C}({I}_1|{A}_1,n),...,\mathbb{C}({I}_{u_i}|{A}_{u_i},n)], 
    \label{eq:purified}
\end{align}

\noindent
where $\mathbb{C}({I}_i|{A}_i,n)$ is the purifying process: 
\begin{align}
\mathbb{C}({I}_i|{A}_i,n)=
\small{\begin{cases}
  \mathbb{C}_d({I}_i \to \emptyset|{A}_i), &  \text{if $I_i$ $\in$ $\mathcal{\hat I}_{u_i}^n$ and $A_i=0$}, \\
  \mathbb{C}_r({I}_i \to {\bar I}_i|{A}_i), &  \text{if $I_i$ $\in$ $\mathcal{\hat I}_{u_i}^n$ and $A_i \ne 0$}, \\
  I_i, &  \text{if $I_i$ $\notin$ $\mathcal{\hat I}_{u_i}^n$}. 
\end{cases}}
\label{eq:whole_purification}
\end{align}

\noindent
By repeating the purification process multiple times on $\mathcal{I}_{u_i}$, we can obtain $m$ cleansed user profiles, where $m$ is a hyperparameter. 
These purified prompts are individually fed into the LLM-empowered RecSys, and the results are subsequently integrated to produce the final recommendation output by using voting mechanisms, defined by:
\begin{align}
    \bar{\textbf{y}} = Voting(\mathcal{R}_{\theta}(\bar{\textbf{x}}_1),\mathcal{R}_{\theta}(\bar{\textbf{x}}_2),...,\mathcal{R}_{\theta}(\bar{\textbf{x}}_m)),
    \label{eq:final_rec}
\end{align}
where $\bar{\textbf{x}}_i=[\mathcal{P} \circ u_i \circ \mathcal{\bar I}_{u_i}]$ is the purified input and $\bar{\textbf{y}}$ is the final recommendation. 
The pseudo-code of RETURN is shown in \textbf{Algorithm}~\ref{al:proposed_method}.

\begin{algorithm}[t]
  \caption{\textbf{RETURN}}  
  \label{al:proposed_method}
  \KwIn{\\
Input $\hat{\textbf{x}}$, External database $\mathcal{E}$, Purification cycle $m$, LLM-empowered RecSys $\mathcal{R}_{\theta}$.\\
\textbf{Output:} Robust recommendations $\bar{\textbf{y}}$.\\
\textbf{Procedure:}}
Generate multi-hop collaborative item graph $\mathcal{G}_{\epsilon}$ according to Eq~\eqref{eq:generate_graph} \;
Retrieve the co-occurrence frequency for a pair of items within $\mathcal{I}_{u_i}$ according to Eq~\eqref{eq:occurrence_score} \;
Compute the occurrence probability for each item within $\mathcal{I}_{u_i}$ according to Eq~\eqref{eq:occurrence_probability} \;
\For{t \text{in} 1:m}
{
Purify the user's historical interactions according to Eq~\eqref{eq:whole_purification} \;
Generate the recommendations for each purified prompt $\bar{\textbf{x}}_i$ and obtain a set of recommendation results $[\mathcal{R}_{\theta}(\bar{\textbf{x}}_1),\mathcal{R}_{\theta}(\bar{\textbf{x}}_2),...,\mathcal{R}_{\theta}(\bar{\textbf{x}}_m)]$ \;
}
Generate the final recommendations based on Eq~\eqref{eq:final_rec} \;
\end{algorithm}  

%% file: Section/Experiments.tex
\section{Experiments}\label{sec:experiments}

\subsection{Experimental Details}

\subsubsection{\textbf{Datasets}. }
All experiments are conducted on three real-world datasets in RecSys: Movielens-1M (\textbf{ML1M})~\cite{movielens}, \textbf{Taobao}~\cite{taobao1}, and \textbf{LastFM}~\cite{xu2024openp5} datasets.  
The ML1M dataset contains one million movie ratings collected from around 6,040 users and their interactions with around 4,000 movies, which is widely used for various recommendation tasks and evaluation of recommendation techniques. 
The LastFM dataset is a widely used music recommendation dataset that contains user listening histories and preferences, which is frequently used to study user preferences, understand music consumption patterns, and evaluate recommendation algorithms. 
The Taobao dataset comprises a massive collection of user interactions on the Taobao e-commerce platform, including browsing, searching, and purchasing activities. It consists of a million records from around 987,994 users and their interactions with around 4,162,024 items and offers valuable insights into user behavior and preferences in the online retail environment. 
For \textbf{P5} model, all the aforementioned datasets are preprocessed following the strategies proposed by~\citet{xu2024openp5}. 
For \textbf{TALLRec} model, it needs to divide the users' historical sequences into users' liked items and disliked items based on their ratings. Since LastFM and Taobao datasets lack rating information from users, we only process the \textbf{ML1M} dataset according to the study of ~\citet{ning2024cheatagent}.

\subsubsection{\textbf{Victim LLM-based Recommender Systems}. }
Two representative LLM-based RecSys, i.e., \textbf{P5} and \textbf{TALLRec}, are employed as the victim models to investigate the performance of different defense techniques. 
\begin{itemize}[leftmargin=*]
    \item \textbf{P5} is a typical ID-based LLM-empowered RecSys, which assigns each item a numerical number and converts the user-item interactions to natural language sequences for recommendations. P5 introduces several item indexing strategies, which can be employed to test the robustness of the defense methods for ID-based RecSys with different indexing strategies.
    \item \textbf{TALLRec} is a representative text-based LLM-empowered recommender system, which integrates textual information (i.e., item title) into a pre-defined prompt template for recommendation. By constructing experiments based on TALLRec, we can investigate the performance of different defense methods for LLM-empowered RecSys employing textual knowledge. 
\end{itemize}

\subsubsection{\textbf{Attackers}. }
We employ CheatAgent~\cite{ning2024cheatagent} as the attacker to generate adversarial perturbations and insert them into the user's historical interactions. 
It should be noted that CheatAgent is an evasion attack method that uses LLMs as the agent to generate high-quality perturbations for misleading the target LLM-empowered RecSys during the inference phase. 
Currently, there is limited research on poisoning attacks for LLM-empowered RecSys. Poisoning attacks require retraining the model, but the large parameter size of LLMs makes frequent retraining infeasible. In other words, poisoning attacks are highly time-consuming for large language models, and they are ineffective if retraining cannot be performed. 
Therefore, in this paper, we solely consider the evasion attack (i.e., CheatAgent) since it is a more efficient attacking method in the era of LLMs. 
We use CheatAgent to generate item perturbations and insert them into the user's history interactions to test the defense performance of different methods.
The primary objective of CheatAgent is to investigate the vulnerabilities of exiting LLM-empowered RecSys, and it allows the insertion of perturbations in both prompt and users' profiles. 
However, during real-world applications, the attacker and users usually have no access to the prompt $\mathcal{P}$, which makes the prompt attack infeasible. 
Therefore, in this paper, we only use CheatAgent to generate item perturbations and insert them into the user's history interactions.

\subsubsection{\textbf{Baselines}. } 
Several baselines are utilized to investigate the defense performance of different methods: 
\begin{itemize}[leftmargin=*]
    \item \textbf{RD}~\cite{zhang2023certified} randomly deletes some items within the users' historical sequences to filter out the adversarial perturbations. 
    \item \textbf{PD}~\cite{jain2023baseline} computes the perplexity for each item and filters out the item with high perplexity for defense. 
    \item \textbf{RPD}~\cite{jain2023baseline} uses an LLM~\cite{chatgpt_paraphraser} to paraphrase the input prompt, which is widely used as the safeguard for LLMs. 
    \item \textbf{RTD}~\cite{jain2023baseline} retokenizes the input prompt, which aims to break tokens apart and disrupt adversarial behaviors. 
    \item \textbf{LLMSI}~\cite{varshney-etal-2024-art} provides a safety instruction, i.e., "Please take into account the noise present in the user's historical interactions and filter them", along with the input prompt to guide the LLM-empowered RecSys to defense adversarial attacks by themselves. 
    \item \textbf{RDE}~\cite{cao2023defending} randomly deletes some items within the users' interaction sequences and generates multiple cleansed prompts. The final recommendations are obtained by majority voting~\cite{lam1997application}.
    \item \textbf{ICL}~\cite{varshney-etal-2024-art} randomly retrieves several users with different historical sequences as the demonstrations and integrates the retrieved users' profiles with the original prompt for recommendation. 
\end{itemize}

\subsubsection{\textbf{Implementation}. }~\label{sec:implementation}
The proposed RETURN and all baselines are implemented by Pytorch. 
All victim models (i.e., \textbf{P5} and \textbf{TALLRec}) and the attacker algorithm (i.e., \textbf{CheatAgent}) are implemented based on their official codes. 
The training and test set are constructed according to the studies of \citet{xu2024openp5} and \citet{bao2023tallrec} for P5 and TALLRec, respectively. 
We adopt CheatAgent to generate adversarial perturbations and insert them into the benign users' interaction history of the test set to investigate the defense performance of different methods. 
The magnitude of perturbations $\triangle$ is set to 3, consistent with the study of \citet{ning2024cheatagent}. 
For the proposed RETURN, we directly use the training set as the external database. 
During the recommendation generation process, $m=10$ is set as default, meaning that the final ensemble recommendation is obtained based on these 10 purified prompts. 
For RD, we randomly delete 3 items and generate recommendations. 
For PD, we select the top 3 items with the highest perplexity as the perturbations and delete these items for recommendations. 
For RTD, we adopt the BPE-dropout~\cite{provilkov-etal-2020-bpe} to tokenize the input query to mitigate the impact of adversarial perturbations. 
RDE generates 10 purified prompts and integrates their recommendation outcomes as the final prediction. 
ICL randomly retrieves 5 users' interaction sequences from the external database and integrates them with the original input for recommendations. 
All random seeds were fixed throughout the experiments, consistent with the used victim RecSys P5~\cite{geng2022recommendation} and TALLRec~\cite{bao2023tallrec}. This ensures that the experimental results are reproducible, and therefore, we do not include variance in the reported results.

\subsubsection{\textbf{Evaluation Metrics}. } 
For \textbf{P5} model, Top-$k$ Hit Ratio (\textbf{H@$k$}) and Normalized Discounted Cumulative Gain (NDCG) (\textbf{N@$k$})~\citep{geng2022recommendation} are employed to evaluate the recommendation performance. In this paper, we set $k=5$ and $k=10$, respectively. 
\textbf{A-H@$k$} and \textbf{A-N@$k$} represent the extent of the decrease in H@$k$ and N@$k$ after inserting adversarial perturbations into the benign prompt,  which are used to measure the attack performance~\cite{ning2024cheatagent}, formulated as: 
\begin{align}
    \text{A-H}@k = 1 - \frac{\widehat{\text{H}@k}}{\text{H}@k}, \text{A-N}@k = 1 - \frac{\widehat{\text{N}@k}}{\text{N}@k},
\end{align}
\noindent where $\widehat{\text{H}}@k$ and $\widehat{\text{N}}@k$ evaluate the recommendation performance of the victim model when it is under attack.
\textbf{D-H@$k$} and \textbf{D-N@$k$} are utilized to evaluate the performance of defense algorithms, which represent the decrease ratio in A-H@$k$ and A-N@$k$, defined as: 
\begin{align}
\text{D-H}@k = \frac{\widetilde{\text{A-H}}@k}{\text{A-H}@k} - 1, \text{D-N}@k = \frac{\widetilde{\text{A-N}}@k}{\text{A-N}@k} - 1,
\end{align}

\noindent where $\widetilde{\text{A-H}}@k$ and $\widetilde{\text{A-N}}@k$ represent the attack performance when adversarial examples are processed by defense algorithms. 
A greater decrease in $\text{A-H}@k$ and $\text{A-N}@k$ indicates reduced attack performance and improved performance of the defense methods.  
For \textbf{TALLRec} model, we utilize the Area Under the Receiver Operating Characteristic (AUC) to assess the recommendation performance, which is consistent with the study of~\citet{bao2023tallrec}.
ASR-A and D-A~\cite{ning2024cheatagent} are employed to evaluate the performance of the attack and defense methods, defined as: 
\begin{align}
\text{ASR-A} = 1-\frac{\widehat{\text{AUC}}}{\text{AUC}}, \text{D-A} = \frac{\widetilde{\text{ASR-A}}}{\text{ASR-A}} - 1,
\end{align}

\noindent where $\widehat{\text{AUC}}$ and $\widetilde{\text{ASR-A}}$ represent the AUC when the input contains perturbations and when the input is purified by the defense methods, respectively.

\begin{table*}[t]
  \centering
  \caption{Defense performance of different methods (Victim model: P5, Indexing: Sequential)}
  \vskip -0.1in
    \scalebox{0.575}{\begin{tabular}{cccccccccccccc}
    \toprule
    \textbf{Datasets} & \textbf{Methods} & \textbf{H@5↑}  & \textbf{H@10↑} & \textbf{N@5↑}  & \textbf{N@10↑} & \textbf{A-H@5↓} & \textbf{A-H@10↓} & \textbf{A-N@5↓} & \textbf{A-N@10↓} & \textbf{D-H@5↑} & \textbf{D-H@10↑} & \textbf{D-N@5↑} & \textbf{D-N@10↑} \\
    \midrule
    \multirow{11}[3]{*}{\textbf{ML1M}} & Benign & 0.2116  & 0.3055  & 0.1436  & 0.1737  & /     & /     & /     & /     & /     & /     & /     & / \\
          & CheatAgent & 0.0646  & 0.1171  & 0.0405  & 0.0573  & 0.6948  & 0.6168  & 0.7181  & 0.6699  & 0.0000  & 0.0000  & 0.0000  & 0.0000  \\
\cmidrule{2-14}          & PD    & 0.1303  & 0.1935  & 0.0851  & 0.1053  & 0.3842  & 0.3664  & 0.4077  & 0.3939  & 0.4471  & 0.4060  & 0.4323  & 0.4120  \\
          & RPD   & 0.0627  & 0.1070  & 0.0389  & 0.0530  & 0.7034  & 0.6499  & 0.7291  & 0.6950  & -0.0124  & -0.0536  & -0.0153  & -0.0374  \\
          & RTD   & 0.0093  & 0.0161  & 0.0060  & 0.0082  & 0.9562  & 0.9474  & 0.9579  & 0.9527  & -0.3761  & -0.5360  & -0.3340  & -0.4222  \\
          & RD    & 0.0969  & 0.1526  & 0.0620  & 0.0799  & 0.5423  & 0.5003  & 0.5680  & 0.5400  & 0.2196  & 0.1889  & 0.2090  & 0.1940  \\
          & LLMSI & 0.0624  & 0.1073  & 0.0398  & 0.0542  & 0.7050  & 0.6488  & 0.7227  & 0.6878  & -0.0146  & -0.0518  & -0.0064  & -0.0267  \\
          & ICL   & 0.0546  & 0.0858  & 0.0348  & 0.0449  & 0.7418  & 0.7192  & 0.7574  & 0.7418  & -0.0676  & -0.1661  & -0.0547  & -0.1073  \\
          & RDE   & 0.0924  & 0.1566  & 0.0581  & 0.0786  & 0.5634  & 0.4873  & 0.5951  & 0.5475  & 0.1892  & 0.2100  & 0.1712  & 0.1827  \\
          & RETURN & \textbf{0.1384} & \textbf{0.2091} & \textbf{0.0915} & \textbf{0.1142} & \textbf{0.3459} & \textbf{0.3154} & \textbf{0.3630} & \textbf{0.3427} & \textbf{0.5023} & \textbf{0.4886} & \textbf{0.4945} & \textbf{0.4885} \\
\hline
    \multirow{11}[2]{*}{\textbf{LastFM}} & Benign & 0.0404  & 0.0606  & 0.0265  & 0.0331  & /     & /     & /     & /     & /     & /     & /     & / \\
          & CheatAgent & 0.0138  & 0.0239  & 0.0084  & 0.0117  & 0.6591  & 0.6061  & 0.6820  & 0.6471  & 0.0000  & 0.0000  & 0.0000  & 0.0000  \\
\cmidrule{2-14}          & PD    & 0.0183  & 0.0330  & 0.0124  & 0.0170  & 0.5455  & 0.4545  & 0.5331  & 0.4851  & 0.1724  & 0.2500  & 0.2183  & 0.2503  \\
          & RPD   & 0.0183  & 0.0312  & 0.0106  & 0.0147  & 0.5455  & 0.4848  & 0.6006  & 0.5556  & 0.1724  & 0.2000  & 0.1193  & 0.1413  \\
          & RTD   & 0.0046  & 0.0110  & 0.0024  & 0.0043  & 0.8864  & 0.8182  & 0.9108  & 0.8691  & -0.3448  & -0.3500  & -0.3355  & -0.3430  \\
          & RD    & 0.0220  & 0.0303  & 0.0139  & 0.0165  & 0.4545  & 0.5000  & 0.4743  & 0.5010  & 0.3103  & 0.1750  & 0.3045  & 0.2258  \\
          & LLMSI & 0.0119  & 0.0229  & 0.0080  & 0.0116  & 0.7045  & 0.6212  & 0.7003  & 0.6491  & -0.0690  & -0.0250  & -0.0269  & -0.0031  \\
          & ICL   & 0.0174  & 0.0321  & 0.0113  & 0.0160  & 0.5682  & 0.4697  & 0.5726  & 0.5172  & 0.1379  & 0.2250  & 0.1604  & 0.2007  \\
          & RDE   & 0.0220  & 0.0339  & 0.0128  & 0.0167  & 0.4545  & 0.4394  & 0.5170  & 0.4960  & 0.3103  & 0.2750  & 0.2420  & 0.2334  \\
          & RETURN & \textbf{0.0266} & \textbf{0.0385} & \textbf{0.0169} & \textbf{0.0207} & \textbf{0.3409} & \textbf{0.3636} & \textbf{0.3613} & \textbf{0.3731} & \textbf{0.4828} & \textbf{0.4000} & \textbf{0.4703} & \textbf{0.4234} \\
\hline
    \multirow{11}[3]{*}{\textbf{Taobao}} & Benign & 0.1420  & 0.1704  & 0.1100  & 0.1191  & /     & /     & /     & /     & /     & /     & /     & / \\
          & CheatAgent & 0.0863  & 0.1099  & 0.0615  & 0.0690  & 0.3922  & 0.3548  & 0.4409  & 0.4207  & 0.0000  & 0.0000  & 0.0000  & 0.0000  \\
\cmidrule{2-14}          & PD    & 0.0935  & 0.1153  & 0.0687  & 0.0758  & 0.3414  & 0.3231  & 0.3752  & 0.3638  & 0.1294  & 0.0894  & 0.1490  & 0.1352  \\
          & RPD   & 0.0811  & 0.1027  & 0.0567  & 0.0637  & 0.4291  & 0.3971  & 0.4845  & 0.4657  & -0.0941  & -0.1192  & -0.0989  & -0.1069  \\
          & RTD   & 0.0016  & 0.0044  & 0.0010  & 0.0019  & 0.9885  & 0.9740  & 0.9908  & 0.9840  & -1.5206  & -1.7453  & -1.2470  & -1.3390  \\
          & RD    & 0.0886  & 0.1121  & 0.0650  & 0.0726  & 0.3760  & 0.3423  & 0.4087  & 0.3905  & 0.0412  & 0.0352  & 0.0731  & 0.0718  \\
          & LLMSI & 0.0867  & 0.1112  & 0.0615  & 0.0694  & 0.3899  & 0.3471  & 0.4408  & 0.4176  & 0.0059  & 0.0217  & 0.0002  & 0.0073  \\
          & ICL   & 0.0557  & 0.0734  & 0.0385  & 0.0442  & 0.6078  & 0.5692  & 0.6502  & 0.6287  & -0.5500  & -0.6043  & -0.4747  & -0.4944  \\
          & RDE   & 0.0855  & 0.1217  & 0.0626  & 0.0742  & 0.3979  & 0.2856  & 0.4305  & 0.3770  & -0.0147  & 0.1951  & 0.0236  & 0.1038  \\
          & RETURN & \textbf{0.1124} & \textbf{0.1384} & \textbf{0.0890} & \textbf{0.0975} & \textbf{0.2088} & \textbf{0.1875} & \textbf{0.1904} & \textbf{0.1817} & \textbf{0.4676} & \textbf{0.4715} & \textbf{0.5682} & \textbf{0.5680} \\
\bottomrule  
\end{tabular}%
}
  \label{tab:p5_sequential_index}%
\end{table*}%

\begin{table*}[t]
  \centering
    \caption{Defense performance of different methods (Victim model: P5, Indexing: Random)}
  \vskip -0.1in
    \scalebox{0.575}{\begin{tabular}{cccccccccccccc}
    \toprule
    \textbf{Datasets} & \textbf{Methods} & \textbf{H@5↑}  & \textbf{H@10↑} & \textbf{N@5↑}  & \textbf{N@10↑} & \textbf{A-H@5↓} & \textbf{A-H@10↓} & \textbf{A-N@5↓} & \textbf{A-N@10↓} & \textbf{D-H@5↑} & \textbf{D-H@10↑} & \textbf{D-N@5↑} & \textbf{D-N@10↑} \\
    \midrule
    \multirow{11}[3]{*}{\textbf{ML1M}} & Benign & 0.1058  & 0.1533  & 0.0693  & 0.0847  & /     & /     & /     & /     & /     & /     & /     & / \\
          & CheatAgent & 0.0421  & 0.0689  & 0.0262  & 0.0348  & 0.6025  & 0.5508  & 0.6221  & 0.5890  & 0.0000  & 0.0000  & 0.0000  & 0.0000  \\
\cmidrule{2-14}          & PD    & 0.0626  & 0.0959  & 0.0407  & 0.0514  & 0.4085  & 0.3747  & 0.4127  & 0.3924  & 0.3221  & 0.3196  & 0.3365  & 0.3338  \\
          & RPD   & 0.0439  & 0.0717  & 0.0280  & 0.0370  & 0.5853  & 0.5324  & 0.5958  & 0.5634  & 0.0286  & 0.0333  & 0.0423  & 0.0435  \\
          & RTD   & 0.0121  & 0.0217  & 0.0076  & 0.0106  & 0.8858  & 0.8585  & 0.8910  & 0.8745  & -0.4701  & -0.5588  & -0.4323  & -0.4847  \\
          & RD    & 0.0425  & 0.0657  & 0.0281  & 0.0355  & 0.5978  & 0.5713  & 0.5950  & 0.5802  & 0.0078  & -0.0373  & 0.0435  & 0.0150  \\
          & LLMSI & 0.0442  & 0.0695  & 0.0271  & 0.0353  & 0.5822  & 0.5464  & 0.6089  & 0.5832  & 0.0338  & 0.0078  & 0.0212  & 0.0099  \\
          & ICL   & 0.0336  & 0.0535  & 0.0214  & 0.0278  & 0.6823  & 0.6512  & 0.6914  & 0.6721  & -0.1325  & -0.1824  & -0.1115  & -0.1411  \\
          & RDE   & 0.0493  & 0.0800  & 0.0303  & 0.0401  & 0.5336  & 0.4784  & 0.5631  & 0.5268  & 0.1143  & 0.1314  & 0.0949  & 0.1057  \\
          & RETURN & \textbf{0.0929} & \textbf{0.1377} & \textbf{0.0604} & \textbf{0.0750} & \textbf{0.1221} & \textbf{0.1015} & \textbf{0.1276} & \textbf{0.1147} & \textbf{0.7974} & \textbf{0.8157} & \textbf{0.7949} & \textbf{0.8053} \\
\hline
    \multirow{11}[2]{*}{\textbf{LastFM}} & Benign & 0.0128  & 0.0248  & 0.0072  & 0.0110  & /     & /     & /     & /     & /     & /     & /     & / \\
          & CheatAgent & 0.0101  & 0.0220  & 0.0055  & 0.0094  & 0.2143  & 0.1111  & 0.2258  & 0.1474  & 0.0000  & 0.0000  & 0.0000  & 0.0000  \\
\cmidrule{2-14}          & PD    & \textbf{0.0174} & \textbf{0.0294} & \textbf{0.0102} & \textbf{0.0139} & \textbf{-0.3571} & \textbf{-0.1852} & \textbf{-0.4227} & \textbf{-0.2609} & \textbf{2.6667} & \textbf{2.6667} & \textbf{2.8719} & \textbf{2.7708} \\
          & RPD   & 0.0128  & 0.0202  & 0.0080  & 0.0104  & 0.0000  & 0.1852  & -0.1203  & 0.0566  & 1.0000  & -0.6667  & 1.5326  & 0.6156  \\
          & RTD   & 0.0128  & 0.0193  & 0.0074  & 0.0095  & 0.0000  & 0.2222  & -0.0392  & 0.1365  & 1.0000  & -1.0000  & 1.1735  & 0.0738  \\
          & RD    & 0.0110  & 0.0229  & 0.0062  & 0.0101  & 0.1429  & 0.0741  & 0.1281  & 0.0808  & 0.3333  & 0.3333  & 0.4327  & 0.4517  \\
          & LLMSI & 0.0101  & 0.0220  & 0.0054  & 0.0093  & 0.2143  & 0.1111  & 0.2451  & 0.1537  & 0.0000  & 0.0000  & -0.0854  & -0.0429  \\
          & ICL   & 0.0073  & 0.0138  & 0.0045  & 0.0066  & 0.4286  & 0.4444  & 0.3746  & 0.4022  & -1.0000  & -3.0000  & -0.6587  & -1.7294  \\
          & RDE   & 0.0110  & 0.0266  & 0.0073  & 0.0123  & 0.1429  & -0.0741  & -0.0161  & -0.1204  & 0.3333  & 1.6667  & 1.0713  & 1.8173  \\
          & RETURN & 0.0138  & 0.0220  & 0.0067  & 0.0093  & -0.0714  & 0.1111  & 0.0692  & 0.1531  & 1.3333  & 0.0000  & 0.6937  & -0.0392  \\
\hline
    \multirow{11}[3]{*}{\textbf{Taobao}} & Benign & 0.1643  & 0.1804  & 0.1277  & 0.1330  & /     & /     & /     & /     & /     & /     & /     & / \\
          & CheatAgent & 0.1012  & 0.1217  & 0.0682  & 0.0749  & 0.3838  & 0.3252  & 0.4661  & 0.4367  & 0.0000  & 0.0000  & 0.0000  & 0.0000  \\
\cmidrule{2-14}          & PD    & 0.1042  & 0.1184  & 0.0725  & 0.0771  & 0.3659  & 0.3433  & 0.4327  & 0.4199  & 0.0468  & -0.0559  & 0.0717  & 0.0384  \\
          & RPD   & 0.0945  & 0.1112  & 0.0640  & 0.0694  & 0.4247  & 0.3833  & 0.4992  & 0.4778  & -0.1065  & -0.1788  & -0.0709  & -0.0942  \\
          & RTD   & 0.0118  & 0.0190  & 0.0073  & 0.0097  & 0.9282  & 0.8946  & 0.9425  & 0.9273  & -1.4182  & -1.7514  & -1.0220  & -1.1234  \\
          & RD    & 0.1094  & 0.1237  & 0.0774  & 0.0820  & 0.3340  & 0.3143  & 0.3941  & 0.3833  & 0.1299  & 0.0335  & 0.1544  & 0.1223  \\
          & LLMSI & 0.1022  & 0.1237  & 0.0684  & 0.0754  & 0.3779  & 0.3143  & 0.4646  & 0.4331  & 0.0156  & 0.0335  & 0.0032  & 0.0082  \\
          & ICL   & 0.0655  & 0.0799  & 0.0465  & 0.0511  & 0.6012  & 0.5568  & 0.6360  & 0.6155  & -0.5662  & -0.7123  & -0.3644  & -0.4094  \\
          & RDE   & 0.1094  & 0.1479  & 0.0793  & 0.0918  & 0.3340  & 0.1798  & 0.3792  & 0.3101  & 0.1299  & 0.4469  & 0.1865  & 0.2899  \\
          & RETURN & \textbf{0.1317} & \textbf{0.1537} & \textbf{0.1055} & \textbf{0.1126} & \textbf{0.1984} & \textbf{0.1480} & \textbf{0.1741} & \textbf{0.1532} & \textbf{0.4831} & \textbf{0.5447} & \textbf{0.6266} & \textbf{0.6491} \\
    \bottomrule
    \end{tabular}%
    }
  \label{tab:p5_randon_index}%
\end{table*}%

\subsection{Defense Effectiveness}
\begin{figure}[t]
\centering
{\includegraphics[width=0.95\linewidth]{{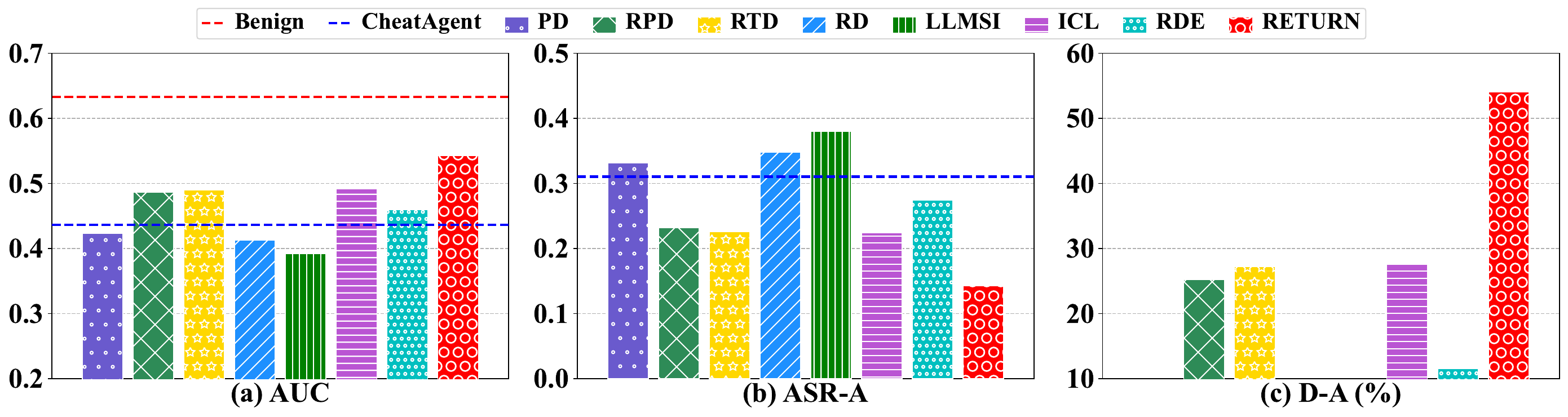}}}
\vskip -0.1in
\caption{Defense Performance on TALLRec}
\label{fig:tallrec_defense}
\end{figure}

In this subsection, we investigate the defense performance of different methods. The results based on \textbf{P5} with different indexing methods are summarised in Table~\ref{tab:p5_sequential_index} and Table~\ref{tab:p5_randon_index}, and the results based on \textbf{TALLRec} are shown in Figure~\ref{fig:tallrec_defense}. 
Benign denotes the use of the original prompt without perturbations for recommendations, and CheatAgent represents the recommendation performance under attacks. 
Based on these experiments, some insights are obtained as follows: 
\begin{itemize}[leftmargin=*]
    \item As shown in Table~\ref{tab:p5_sequential_index} and Table~\ref{tab:p5_randon_index}, the recommendation performance increases after deleting high perplexity items using PD. 
    However, the effectiveness of this method is not robust. For instance, on the ML1M dataset, PD can significantly enhance the recommendation performance of RecSys under attacks. While on the Taobao dataset, the defense performance of PD is limited. 
    
    \item RPD and RTD, two common defense methods for LLMs, cannot achieve the desired performance for LLM-empowered RecSys in most cases. 
    The reason is that LLM-empowered RecSys have captured the domain-specific knowledge of recommendations (e.g., the meaning of item IDs and item relationships) during the training process. 
    However, the LLMs employed by RPD struggle to understand item IDs, making it challenging to effectively rewrite the input prompt. 
    Additionally, RTD disrupts the item ID structure, which further degrades recommendation performance. 

    \item Adversarial perturbations are typically carefully crafted, so disrupting any component may reduce the attack's effectiveness. Therefore, randomly removing a few items from the user's interaction history (i.e., RD) can improve the robustness of the LLM-powered RecSys. Furthermore, RDE generally outperforms RD, suggesting that an ensemble strategy can further enhance system robustness.

    \item The proposed RETURN outperforms all other baselines on three datasets and significantly improves the recommendation performance even under attacks, demonstrating the potential of introducing collaborative knowledge from external databases. For example, on the Taobao dataset, CheatAgent reduces the H@5 from 0.1420 to 0.0863. By introducing collaborative knowledge for input purification, RETURN raises the H@5 to 0.1124, nearly approaching the recommendation performance of using benign prompts, which fully demonstrates the effectiveness of RETURN. 

    \item TALLRec uses item titles to construct the input prompt, which has distinct inherent mechanisms with P5. As shown in Figure~\ref{fig:tallrec_defense}, the proposed RETURN also dramatically increases the AUC of TALLRec and decreases the attack performance, demonstrating the robustness of RETURN to the architecture of the LLM-empowered RecSys. 
\end{itemize}

\begin{table*}[t]
  \centering
    \caption{The defense performance of the proposed RETURN with respect to different attack methods}
    \vskip -0.1in
    \scalebox{0.625}{\begin{tabular}{ccccccccccccc}
    \toprule
    \textbf{Methods} & \textbf{H@5↑}  & \textbf{H@10↑} & \textbf{N@5↑}  & \textbf{N@10↑} & \textbf{A-H@5↓} & \textbf{A-H@10↓} & \textbf{A-N@5↓} & \textbf{A-N@10↓} & \textbf{D-H@5↑} & \textbf{D-H@10↑} & \textbf{D-N@5↑} & \textbf{D-N@10↑} \\
    \midrule
    Benign & 0.0404  & 0.0606  & 0.0265  & 0.0331  & /     & /     & /     & /     & /     & /     & /     & / \\
    \midrule
    PA    & 0.0064  & 0.0147  & 0.0032  & 0.0060  & 0.8409  & 0.7576  & 0.8777  & 0.8187  & 0.0000  & 0.0000  & 0.0000  & 0.0000  \\
    RETURN & 0.0248  & 0.0376  & 0.0148  & 0.0189  & 0.3864  & 0.3788  & 0.4423  & 0.4280  & 0.5405  & 0.5000  & 0.4961  & 0.4772  \\
    \midrule
    RA    & 0.0376  & 0.0587  & 0.0251  & 0.0317  & 0.0682  & 0.0303  & 0.0540  & 0.0405  & 0.0000  & 0.0000  & 0.0000  & 0.0000  \\
    RETURN & 0.0394  & 0.0587  & 0.0257  & 0.0318  & 0.0227  & 0.0303  & 0.0326  & 0.0377  & 0.6667  & 0.0000  & 0.3957  & 0.0679  \\
    \midrule
    CheatAgent & 0.0138  & 0.0239  & 0.0084  & 0.0117  & 0.6591  & 0.6061  & 0.6820  & 0.6471  & 0.0000  & 0.0000  & 0.0000  & 0.0000  \\
    RETURN & 0.0266  & 0.0385  & 0.0169  & 0.0207  & 0.3409  & 0.3636  & 0.3613  & 0.3731  & 0.4828  & 0.4000  & 0.4703  & 0.4234  \\
    \bottomrule
    \end{tabular}%
    }
  \label{tab:perturbation_methods}%
\end{table*}%

\begin{table*}[t]
  \centering
  \caption{Comparison between RETURN and its variants on three datasets }
    \vskip -0.1in
    \scalebox{0.575}{\begin{tabular}{clcccccccccccc}
    \toprule
    \textbf{Datasets} & \textbf{Methods} & \textbf{H@5↑}  & \textbf{H@10↑} & \textbf{N@5↑}  & \textbf{N@10↑} & \textbf{A-H@5↓} & \textbf{A-H@10↓} & \textbf{A-N@5↓} & \textbf{A-N@10↓} & \textbf{D-H@5↑} & \textbf{D-H@10↑} & \textbf{D-N@5↑} & \textbf{D-N@10↑} \\
    \midrule
    \multirow{6}[3]{*}{\textbf{ML1M}} & Benign & 0.2116  & 0.3055  & 0.1436  & 0.1737  & /     & /     & /     & /     &       &       &       &  \\
          & CheatAgent & 0.0646  & 0.1171  & 0.0405  & 0.0573  & 0.6948  & 0.6168  & 0.7181  & 0.6699  & 0.0000  & 0.0000  & 0.0000  & 0.0000  \\
\cmidrule{2-14}          & RETURN & \textbf{0.1384 } & \textbf{0.2091 } & \textbf{0.0915 } & \textbf{0.1142 } & \textbf{0.3459 } & \textbf{0.3154 } & \textbf{0.3630 } & \textbf{0.3427 } & \textbf{0.5023 } & \textbf{0.4886 } & \textbf{0.4945 } & \textbf{0.4885 } \\
          & ~~~--ROP   & 0.0747  & 0.1286  & 0.0467  & 0.0639  & 0.6471  & 0.5789  & 0.6747  & 0.6321  & 0.0687  & 0.0615  & 0.0604  & 0.0564  \\
          & ~~~--RR    & 0.1093  & 0.1705  & 0.0701  & 0.0898  & 0.4836  & 0.4417  & 0.5119  & 0.4831  & 0.3041  & 0.2838  & 0.2872  & 0.2788  \\
          & ~~~--w/o Ens & 0.1185  & 0.1889  & 0.0783  & 0.1010  & 0.4397  & 0.3816  & 0.4546  & 0.4184  & 0.3671  & 0.3814  & 0.3670  & 0.3754  \\ 
\hline 
          \multirow{6}[2]{*}{\textbf{LastFM}} & Benign & 0.0404  & 0.0606  & 0.0265  & 0.0331  & /     & /     & /     & /     &       &       &       &  \\
          & CheatAgent & 0.0138  & 0.0239  & 0.0084  & 0.0117  & 0.6591  & 0.6061  & 0.6820  & 0.6471  & 0.0000  & 0.0000  & 0.0000  & 0.0000  \\
\cmidrule{2-14}          & RETURN & \textbf{0.0266 } & \textbf{0.0385 } & \textbf{0.0169 } & \textbf{0.0207 } & \textbf{0.3409 } & \textbf{0.3636 } & \textbf{0.3613 } & \textbf{0.3731 } & \textbf{0.4828 } & \textbf{0.4000 } & \textbf{0.4703 } & \textbf{0.4234 } \\
          & ~~~--ROP   & 0.0165  & 0.0321  & 0.0115  & 0.0164  & 0.5909  & 0.4697  & 0.5683  & 0.5044  & 0.1034  & 0.2250  & 0.1667  & 0.2205  \\
          & ~~~--RR    & 0.0248  & 0.0339  & 0.0147  & 0.0176  & 0.3864  & 0.4394  & 0.4466  & 0.4674  & 0.4138  & 0.2750  & 0.3452  & 0.2778  \\
          & ~~~--w/o Ens & 0.0248  & 0.0367  & 0.0151  & 0.0188  & 0.3864  & 0.3939  & 0.4304  & 0.4309  & 0.4138  & 0.3500  & 0.3689  & 0.3342  \\
\hline 
    \multirow{6}[3]{*}{\textbf{Taobao}} & Benign & 0.1420  & 0.1704  & 0.1100  & 0.1191  & /     & /     & /     & /     &       &       &       &  \\
          & CheatAgent & 0.0863  & 0.1099  & 0.0615  & 0.0690  & 0.3922  & 0.3548  & 0.4409  & 0.4207  & 0.0000  & 0.0000  & 0.0000  & 0.0000  \\
\cmidrule{2-14}          & RETURN & \textbf{0.1124 } & \textbf{0.1384 } & \textbf{0.0890 } & \textbf{0.0975 } & \textbf{0.2088 } & \textbf{0.1875 } & \textbf{0.1904 } & \textbf{0.1817 } & \textbf{0.4676 } & \textbf{0.4715 } & \textbf{0.5682 } & \textbf{0.5680 } \\
          & ~~~--ROP   & 0.1008  & 0.1222  & 0.0749  & 0.0819  & 0.2907  & 0.2827  & 0.3185  & 0.3126  & 0.2588  & 0.2033  & 0.2776  & 0.2570  \\
          & ~~~--RR    & 0.1122  & 0.1376  & 0.0882  & 0.0964  & 0.2099  & 0.1923  & 0.1981  & 0.1911  & 0.4647  & 0.4580  & 0.5508  & 0.5457  \\
          & ~~~--w/o Ens & 0.1006  & 0.1250  & 0.0765  & 0.0843  & 0.2918  & 0.2663  & 0.3046  & 0.2925  & 0.2559  & 0.2493  & 0.3093  & 0.3048  \\
    \bottomrule
    \end{tabular}%
    }
  \label{tab:ablation}%
\end{table*}%

\subsection{Model Analysis}

\subsubsection{Attack Scenarios: Clickbait and vulnerabilities of LLM-empowered RecSys.}
The attack discussed in this paper mirrors a real-world phenomenon, commonly known as clickbait~\cite{yue2010beyond,wang2021clicks}. Clickbait refers to the scenario in which attackers might post products with enticing images and titles to attract user clicks on an e-commerce platform. Users are easily drawn to these clickbait products and interact with them, even though the content of these goods may not truly align with their preferences~\cite{yue2010beyond,wang2021clicks}. However, existing studies~\cite{ning2024cheatagent} have demonstrated that LLM-empowered RecSys is vulnerable to minor perturbations in user historical interactions. If users are attracted by clickbait products and engage with them, minor perturbations will be introduced to their historical interactions. Such minor perturbations (e.g., irrelevant items) can easily lead the LLM-empowered RecSys to misunderstand the user preferences by capturing the collaborative knowledge from the user’s historical interactions. This leads to inaccurate recommendations, affecting user experience and engagement and consequently diminishing company profits. Therefore, enhancing the robustness of the LLM-empowered RecSys is crucial to mitigate the clickbait issue, which is a practical necessity. 

During experiments, to simulate the worst-case scenario, we adopt CheatAgent~\cite{ning2024cheatagent}, which is a powerful attacker, to insert perturbations to the user's historical sequences. Besides, we also employ various attack methods and perturbation intensities to simulate the scenario in which the user's historical interactions contain minor perturbations. We adopt two other methods to generate adversarial perturbations: PA~\cite{xullm} adopts an LLM to generate perturbations, and RA~\cite{ning2024cheatagent} randomly selects the items from the item pool as the perturbations. 

As shown in Table~\ref{tab:perturbation_methods}, we can observe that the proposed defense method significantly reduces the effectiveness of various attack methods (i.e., CheatAgent, PA). This implies that even if users interact with clickbait items that trigger vulnerabilities in the recommendation system, the proposed RETURN method can effectively cleanse these malicious disturbances, ensuring the correctness of recommendations. Regarding RA, its attack capability is constrained, and it is aimed at simulating scenarios where perturbation items do not cause the RecSys to misinterpret user preferences. In this case, RETURN still improves or maintains the recommendation performance of the RecSys. This demonstrates the robustness of the proposed RETURN against different attack intensities and scenarios.

\subsubsection{Ablation Study}
Three variants \textbf{RETURN-ROP}, \textbf{RETURN-RR}, and \textbf{RETURN-w/o Ens} are employed for comparison: 
1) \textbf{RETURN-ROP} randomly creates the collaborative item graphs to demonstrate the effectiveness and importance of introducing the external database. 
2) \textbf{RETURN-RR} directly deletes all items with low occurrence probabilities. 
3) \textbf{RETURN-w/o Ens} generates recommendations without using the ensemble strategy and only creates one purified prompt by processing a fixed number of items. 
The results are summarised in Table~\ref{tab:ablation}. 
RETURN-ROP generates recommendations without constructing collaborative item graphs from the external database, resulting in a significant decrease in its defense performance. This highlights the importance of introducing accurate collaborative knowledge from the external database. 
Since directly deleting all items with low occurrence probabilities may result in the RecSys failing to capture users' preferences effectively, especially for users with limited interactions, there is a significant decrease in the defense performance of RETURN-RR, illustrating the importance of employing the retrieval-augmented denoising strategy. 
Since the number of the perturbations is unknown, RETURN-w/o Ens fixes the number of purification items. This approach usually leads to information loss if an excessive number of items are deleted, or incomplete purification if not all perturbations are eliminated, demonstrating the importance of the robust ensemble recommendation strategy.

\subsubsection{Parameter Analysis}
\begin{figure}[t]
\centering
\subfigure[$\text{H}@k$ and $\text{N}@k$ w.r.t. $m$]{
	\includegraphics[width=1.8in]{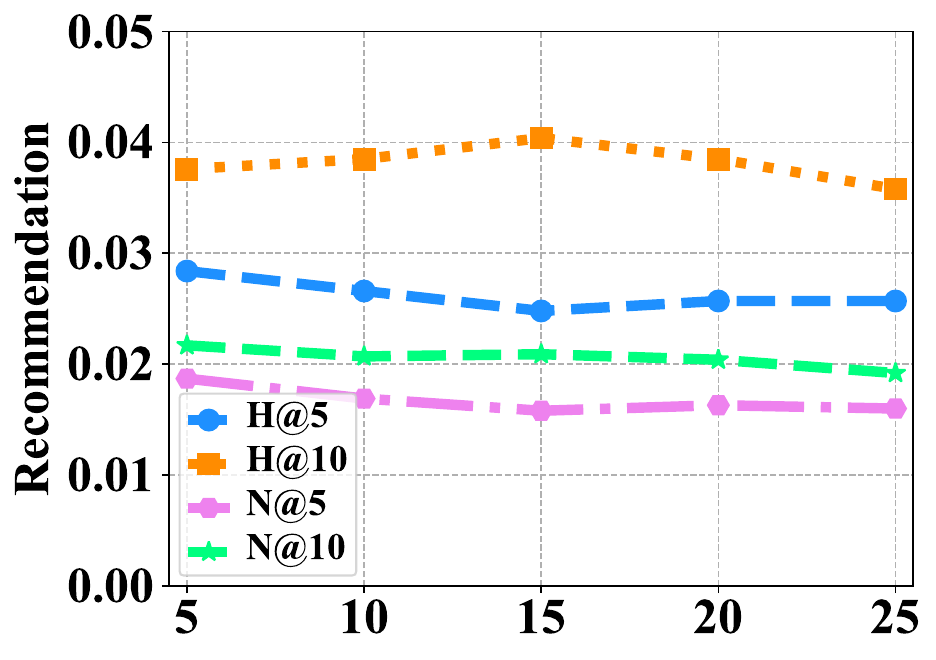}
}
\subfigure[$\text{D-H}@k$ and $\text{D-N}@k$ w.r.t. $m$]{
	\includegraphics[width=1.775in]{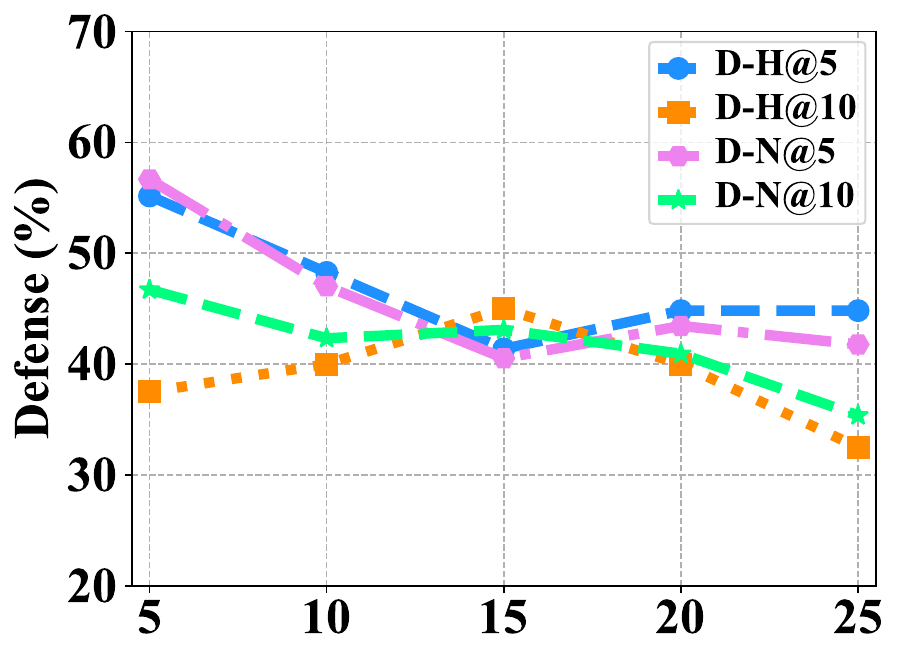}
}
\caption{Effect of the hyper-parameters $m$. }
\label{fig:parameter_analysis}
\end{figure}

We investigate the sensitivity of \method to the hyperparameter $m$. We sample varying values for $m$ and test the defense performance of the proposed method. and the results are illustrated in Figure~\ref{fig:parameter_analysis}. 
We observe that as $m$ increases, the recommendation performance and the defense capability of \method fluctuate within a small range, demonstrating the robustness of the proposed method to hyperparameters.

\subsubsection{The Robustness to the Perturbation Intensity}\label{appendix:perturbation_intensity}
In this subsection, we investigate the robustness of RETURN to the perturbation intensity $\triangle$. 
We insert varying numbers of perturbations into benign users and evaluate the defense performance of the proposed method. 
As shown in Table~\ref{tab:perturbation_intensity}, the proposed method significantly enhances the recommendation performance of LLM-empowered RecSys regardless of the number of perturbations inserted into the input. 
This is attributed to robust recommendation generation strategies that avoid introducing fixed thresholds, thereby improving the robustness of the proposed RETURN to the number of perturbations.

\begin{table*}[t]
  \centering
    \caption{The defense performance of RETURN with respect to the perturbation intensity $\triangle$}
      \vskip -0.1in
    \scalebox{0.625}{\begin{tabular}{ccccccccccccc}
    \toprule
    \textbf{Methods} & \textbf{H@5↑}  & \textbf{H@10↑} & \textbf{N@5↑}  & \textbf{N@10↑} & \textbf{A-H@5↓} & \textbf{A-H@10↓} & \textbf{A-N@5↓} & \textbf{A-N@10↓} & \textbf{D-H@5↑} & \textbf{D-H@10↑} & \textbf{D-N@5↑} & \textbf{D-N@10↑} \\
    \midrule
    Benign & 0.0404  & 0.0606  & 0.0265  & 0.0331  & /     & /     & /     & /     & /     & /     & /     & / \\
    \midrule
    $\triangle$=1 & 0.0183  & 0.0394  & 0.0122  & 0.0190  & 0.5455  & 0.3485  & 0.5390  & 0.4265  & 0.0000  & 0.0000  & 0.0000  & 0.0000  \\
    RETURN & 0.0303  & 0.0486  & 0.0208  & 0.0265  & 0.2500  & 0.1970  & 0.2149  & 0.1974  & 0.5417  & 0.4348  & 0.6013  & 0.5373  \\
    \midrule
    $\triangle$=2 & 0.0138  & 0.0257  & 0.0090  & 0.0128  & 0.6591  & 0.5758  & 0.6621  & 0.6120  & 0.0000  & 0.0000  & 0.0000  & 0.0000  \\
    RETURN & 0.0248  & 0.0367  & 0.0148  & 0.0185  & 0.3864  & 0.3939  & 0.4432  & 0.4406  & 0.4138  & 0.3158  & 0.3307  & 0.2801  \\
    \midrule
    $\triangle$=3 & 0.0138  & 0.0239  & 0.0084  & 0.0117  & 0.6591  & 0.6061  & 0.6820  & 0.6471  & 0.0000  & 0.0000  & 0.0000  & 0.0000  \\
    RETURN & 0.0266  & 0.0385  & 0.0169  & 0.0207  & 0.3409  & 0.3636  & 0.3613  & 0.3731  & 0.4828  & 0.4000  & 0.4703  & 0.4234  \\
    \midrule
    $\triangle$=4 & 0.0119  & 0.0202  & 0.0082  & 0.0109  & 0.7045  & 0.6667  & 0.6893  & 0.6704  & 0.0000  & 0.0000  & 0.0000  & 0.0000  \\
    RETURN & 0.0248  & 0.0349  & 0.0162  & 0.0194  & 0.3864  & 0.4242  & 0.3879  & 0.4146  & 0.4516  & 0.3636  & 0.4372  & 0.3816  \\
    \midrule
    $\triangle$=5 & 0.0119  & 0.0211  & 0.0068  & 0.0098  & 0.7045  & 0.6515  & 0.7425  & 0.7048  & 0.0000  & 0.0000  & 0.0000  & 0.0000  \\
    RETURN & 0.0211  & 0.0284  & 0.0149  & 0.0173  & 0.4773  & 0.5303  & 0.4385  & 0.4771  & 0.3226  & 0.1860  & 0.4094  & 0.3231  \\
    \bottomrule
    \end{tabular}%
    }
  \label{tab:perturbation_intensity}%
\end{table*}%

\subsubsection{Impact on Benign Users}\label{appendix:impact_to_benign}

It is crucial that defense algorithms should not affect the recommendation performance of RecSys for users whose interaction histories contain no perturbations. 
Therefore, in this subsection, the impact of RETURN on benign users is investigated, and the results are shown in Table~\ref{tab:defense_benign}. 
We can observe that if the users' profiles consist of no perturbations, RETURN can almost maintain the recommendation performance even though RETURN deletes or replaces some items. 
Note that the deletion or replacement operations are implemented based on the collaborative co-occurrence frequency, indicating that the selected items usually fail to align with the users' preferences. 
Therefore, RETURN has little impact on the recommendation effectiveness for benign users, which demonstrates its practical applicability in enhancing the robustness of LLM-empowered RecSys.

\begin{table}[htbp]
  \centering
  \caption{Recommendation performance when users' profiles contain no perturbation}
  \scalebox{0.75}{
    \begin{tabular}{ccccccc}
    \toprule
    \textbf{Indexing} & \textbf{Datasets} & \textbf{Methods} & \textbf{H@5↑}  & \textbf{H@10↑} & \textbf{N@5↑}  & \textbf{N@10↑} \\
    \midrule
    \multirow{6}[6]{*}{\textbf{Sequential}} & \multirow{2}[2]{*}{ML1M} & Benign & 0.2116  & 0.3055  & 0.1436  & 0.1737  \\
          &       & RETURN & 0.1675  & 0.2498  & 0.1131  & 0.1397  \\
\cmidrule{2-7}          & \multirow{2}[2]{*}{LastFM} & Benign & 0.0404  & 0.0606  & 0.0265  & 0.0331  \\
          &       & RETURN & 0.0376  & 0.0569  & 0.0232  & 0.0293  \\
\cmidrule{2-7}          & \multirow{2}[2]{*}{Taobao} & Benign & 0.1420  & 0.1704  & 0.1100  & 0.1191  \\
          &       & RETURN & 0.1006  & 0.1250  & 0.0765  & 0.0843  \\
    \midrule
    \multirow{6}[6]{*}{\textbf{Random}} & \multirow{2}[2]{*}{ML1M} & Benign & 0.1058  & 0.1533  & 0.0693  & 0.0847  \\
          &       & RETURN & 0.0944  & 0.1406  & 0.0611  & 0.0760  \\
\cmidrule{2-7}          & \multirow{2}[2]{*}{LastFM} & Benign & 0.0128  & 0.0248  & 0.0072  & 0.0110  \\
          &       & RETURN & 0.0156  & 0.0284  & 0.0094  & 0.0136  \\
\cmidrule{2-7}          & \multirow{2}[2]{*}{Taobao} & Benign & 0.1643  & 0.1804  & 0.1277  & 0.1330  \\
          &       & RETURN & 0.1239  & 0.1409  & 0.0893  & 0.0948  \\
    \bottomrule
    \end{tabular}%
    }
  \label{tab:defense_benign}%
\end{table}%

\subsubsection{Time Complexity}\label{appendix:time_complexity}
To address the concern regarding the computational overhead introduced by the RETURN framework, we conduct additional experiments to analyse the time complexity of RETURN. We measure the average time taken by the LLM-empowered RecSys to generate recommendations after incorporating different defense methods on the LastFM dataset. As shown in Tabel~\ref{tab:time_complexity}, we can observe that methods requiring minimal computational resources (e.g., RD, LLMSI, etc.) exhibit significantly shorter recommendation generation times, typically less than 0.5 seconds. However, their defense performance is notably limited. In contrast, more powerful methods, including RETURN, exhibit slightly longer recommendation generation times, with RETURN taking approximately 0.8599 seconds. This is comparable to other advanced defense methods like PD (0.7314 seconds) and RDE (0.7222 seconds), which also take around 1 second. 

The results indicate that while RETURN introduces additional computational steps, such as voting operations, it does not significantly increase the overall computational burden of the RecSys. Importantly, RETURN achieves this while substantially enhancing the robustness of RecSys against perturbations. Thus, the framework strikes a balance between computational efficiency and defense effectiveness, making it a practical choice for real-world applications.

\begin{table}[htbp]
  \centering
    \caption{Computational time of different methods}
    \scalebox{0.8}{\begin{tabular}{ccccccccc}
    \toprule
    \textbf{Methods} & \textbf{PD}    & \textbf{RPD}   & \textbf{RTD}   & \textbf{RD}    & \textbf{LLMSI} & \textbf{ICL}   & \textbf{RDE}   & \textbf{RETURN} \\
    \midrule
    \textbf{Time (s)}  & 0.7314  & 1.2271  & 0.2916  & 0.3036  & 0.3006  & 0.3130  & 0.7222  & 0.8599  \\
    \bottomrule
    \end{tabular}%
    }
  \label{tab:time_complexity}%
\end{table}%

\subsubsection{Impact of Poor Quality Data}
We conduct additional experiments to investigate the impact of data quality. We introduce two variants: \textbf{RETURN-A-k} and \textbf{RETURN-D-k}, where perturbations are injected into or items are deleted from the historical interactions of users in the external database to generate collaborative item graphs. Here, $k$=0.15 and $k$=0.3 represent the proportion of perturbations or deletions, respectively. The results are shown in Table~\ref{tab:data_quality}. 
The results demonstrate that RETURN-A-k still achieves remarkable defense performance even when perturbations are introduced into the external database. This is because the collaborative item graphs store co-occurrence frequencies, and minor perturbations do not significantly alter the overall co-occurrence distribution among items. After normalization, these perturbations have minimal impact on RETURN's ability to cleanse user interaction data and generate accurate recommendations.
Additionally, RETURN-D-k fails to achieve the desired defense performance because the lack of sufficient collaborative signals prevents it from accurately capturing relationships between items, thereby hindering its ability to identify perturbations. 

These experimental results indicate that the presence of noisy data in the external database (i.e., low-quality data) does not significantly deteriorate the performance of RETURN, as the co-occurrence distribution remains relatively stable. However, insufficient data (e.g., due to deletions) can degrade RETURN's defense effectiveness, as it relies on sufficient collaborative signals to accurately model item relationships. Therefore, while RETURN is robust to minor data quality issues, ensuring an adequate volume of data is crucial for maintaining its performance.

\begin{table*}[htbp]
  \centering
  \caption{The defense performance of RETURN with respect to different external databases }
  \vskip -0.1in
    \scalebox{0.6}{\begin{tabular}{ccccccccccccc}
    \toprule
    \textbf{Methods} & \textbf{H@5↑}  & \textbf{H@10↑} & \textbf{N@5↑}  & \textbf{N@10↑} & \textbf{A-H@5↓} & \textbf{A-H@10↓} & \textbf{A-N@5↓} & \textbf{A-N@10↓} & \textbf{D-H@5↑} & \textbf{D-H@10↑} & \textbf{D-N@5↑} & \textbf{D-N@10↑} \\
    \midrule
    Benign & 0.0404  & 0.0606  & 0.0265  & 0.0331  & /     & /     & /     & /     & /     & /     & /     & / \\
    CheatAgent & 0.0138  & 0.0239  & 0.0084  & 0.0117  & 0.6591  & 0.6061  & 0.6820  & 0.6471  & 0.0000  & 0.0000  & 0.0000  & 0.0000  \\
    \midrule
    PD    & 0.0183  & 0.0330  & 0.0124  & 0.0170  & 0.5455  & 0.4545  & 0.5331  & 0.4851  & 0.1724  & 0.2500  & 0.2183  & 0.2503  \\
    RPD   & 0.0183  & 0.0312  & 0.0106  & 0.0147  & 0.5455  & 0.4848  & 0.6006  & 0.5556  & 0.1724  & 0.2000  & 0.1193  & 0.1413  \\
    RTD   & 0.0046  & 0.0110  & 0.0024  & 0.0043  & 0.8864  & 0.8182  & 0.9108  & 0.8691  & -0.3448  & -0.3500  & -0.3355  & -0.3430  \\
    RD    & 0.0220  & 0.0303  & 0.0139  & 0.0165  & 0.4545  & 0.5000  & 0.4743  & 0.5010  & 0.3103  & 0.1750  & 0.3045  & 0.2258  \\
    LLMSI & 0.0119  & 0.0229  & 0.0080  & 0.0116  & 0.7045  & 0.6212  & 0.7003  & 0.6491  & -0.0690  & -0.0250  & -0.0269  & -0.0031  \\
    ICL   & 0.0174  & 0.0321  & 0.0113  & 0.0160  & 0.5682  & 0.4697  & 0.5726  & 0.5172  & 0.1379  & 0.2250  & 0.1604  & 0.2007  \\
    RDE   & 0.0220  & 0.0339  & 0.0128  & 0.0167  & 0.4545  & 0.4394  & 0.5170  & 0.4960  & 0.3103  & 0.2750  & 0.2420  & 0.2334  \\
    \midrule
    RETURN & 0.0266  & 0.0385  & 0.0169  & 0.0207  & 0.3409  & 0.3636  & 0.3613  & 0.3731  & 0.4828  & 0.4000  & 0.4703  & 0.4234  \\
    RETURN - A - 0.15 & 0.0294  & 0.0413  & 0.0180  & 0.0219  & 0.2727  & 0.3182  & 0.3208  & 0.3391  & 0.5862  & 0.4750  & 0.5296  & 0.4760  \\
    RETURN - A - 0.3 & 0.0294  & 0.0440  & 0.0185  & 0.0232  & 0.2727  & 0.2727  & 0.3020  & 0.2980  & 0.5862  & 0.5500  & 0.5571  & 0.5395  \\
    RETURN - D - 0.15 & 0.0165  & 0.0321  & 0.0103  & 0.0152  & 0.5909  & 0.4697  & 0.6125  & 0.5409  & 0.1034  & 0.2250  & 0.1019  & 0.1642  \\
    RETURN - D - 0.3 & 0.0165  & 0.0321  & 0.0103  & 0.0152  & 0.5909  & 0.4697  & 0.6125  & 0.5409  & 0.1034  & 0.2250  & 0.1019  & 0.1642  \\
    \bottomrule
    \end{tabular}%
    }
  \label{tab:data_quality}%
\end{table*}%

\subsubsection{The Adoption of Normal Distribution}\label{appendix:normal_distribution}
During the robust ensemble recommendation process, RETURN randomly samples an integer $n$ from a normal distribution, and Top-$n$ items with the lowest occurrence probabilities are identified from the user's historical interactions for purification. 
The normal distribution is chosen because it allows for better control over the strength of perturbation filtering in RETURN. If a majority of users' interaction histories contain significant perturbations, making it difficult for RecSys to accurately capture their preferences, the mean can be adjusted to enhance the purification strength of RETURN. 

During experiments, the mean and the variance are 3.5 and 0.5, respectively. 
Moreover, we conducted additional experiments to demonstrate that RETURN is robust to the different values of mean and variance. The results are shown in Table~\ref{tab:normal_distribution}. 
The performance of RETURN fluctuates within a reasonable range as the mean and variance change, demonstrating its robustness to different parameter settings. This indicates that RETURN can adapt to varying distributions while maintaining its effectiveness in generating accurate recommendations.

\begin{table*}[htbp]
  \centering
  \caption{The defense performance of RETURN with respect to different values of mean and variance }
  \vskip -0.1in
    \scalebox{0.625}{\begin{tabular}{ccccccccccccc}
    \toprule
    \textbf{Methods} & \textbf{H@5↑}  & \textbf{H@10↑} & \textbf{N@5↑}  & \textbf{N@10↑} & \textbf{A-H@5↓} & \textbf{A-H@10↓} & \textbf{A-N@5↓} & \textbf{A-N@10↓} & \textbf{D-H@5↑} & \textbf{D-H@10↑} & \textbf{D-N@5↑} & \textbf{D-N@10↑} \\
    \midrule
    Benign & 0.0404  & 0.0606  & 0.0265  & 0.0331  & /     & /     & /     & /     &       &       &       &  \\
    CheatAgent & 0.0138  & 0.0239  & 0.0084  & 0.0117  & 0.6591  & 0.6061  & 0.6820  & 0.6471  & 0.0000  & 0.0000  & 0.0000  & 0.0000  \\
    \midrule
    PD    & 0.0183  & 0.0330  & 0.0124  & 0.0170  & 0.5455  & 0.4545  & 0.5331  & 0.4851  & 0.1724  & 0.2500  & 0.2183  & 0.2503  \\
    RPD   & 0.0183  & 0.0312  & 0.0106  & 0.0147  & 0.5455  & 0.4848  & 0.6006  & 0.5556  & 0.1724  & 0.2000  & 0.1193  & 0.1413  \\
    RTD   & 0.0046  & 0.0110  & 0.0024  & 0.0043  & 0.8864  & 0.8182  & 0.9108  & 0.8691  & -0.3448  & -0.3500  & -0.3355  & -0.3430  \\
    RD    & 0.0220  & 0.0303  & 0.0139  & 0.0165  & 0.4545  & 0.5000  & 0.4743  & 0.5010  & 0.3103  & 0.1750  & 0.3045  & 0.2258  \\
    LLMSI & 0.0119  & 0.0229  & 0.0080  & 0.0116  & 0.7045  & 0.6212  & 0.7003  & 0.6491  & -0.0690  & -0.0250  & -0.0269  & -0.0031  \\
    ICL   & 0.0174  & 0.0321  & 0.0113  & 0.0160  & 0.5682  & 0.4697  & 0.5726  & 0.5172  & 0.1379  & 0.2250  & 0.1604  & 0.2007  \\
    RDE   & 0.0220  & 0.0339  & 0.0128  & 0.0167  & 0.4545  & 0.4394  & 0.5170  & 0.4960  & 0.3103  & 0.2750  & 0.2420  & 0.2334  \\
    \midrule
    N(3.5, 0.5) & 0.0266  & 0.0385  & 0.0169  & 0.0207  & 0.3409  & 0.3636  & 0.3613  & 0.3731  & 0.4828  & 0.4000  & 0.4703  & 0.4234  \\
    N(3, 0.5) & 0.0229  & 0.0358  & 0.0132  & 0.0173  & 0.4318  & 0.4091  & 0.5035  & 0.4766  & 0.3448  & 0.3250  & 0.2618  & 0.2635  \\
    N(4, 0.5) & 0.0267  & 0.0413  & 0.0174  & 0.0224  & 0.3389  & 0.3182  & 0.3449  & 0.3236  & 0.4859  & 0.4750  & 0.4943  & 0.4999  \\
    N(3.5, 1.0) & 0.0229  & 0.0376  & 0.0150  & 0.0197  & 0.4318  & 0.3788  & 0.4362  & 0.4043  & 0.3448  & 0.3750  & 0.3604  & 0.3752  \\
    N(3.5, 1.5) & 0.0220  & 0.0367  & 0.0150  & 0.0198  & 0.4545  & 0.3939  & 0.4349  & 0.4025  & 0.3103  & 0.3500  & 0.3624  & 0.3780  \\
    \bottomrule
    \end{tabular}%
    }
  \label{tab:normal_distribution}%
\end{table*}%

\subsubsection{Impact on the Personalization of Recommendations}
To evaluate the impact of RETURN on personalized recommendations, we separately analyze the recommendation results of the LLM-empowered RecSys for benign users and the results after introducing RETURN for denoising. We calculate the frequency of different items in both sets of results, computed the Jaccard similarity coefficient~\cite{bag2019efficient} between the two distributions, determined the proportion of items that co-occurred, and measured the Shannon entropy of each distribution. The results are presented in Table~\ref{tab:impact_to_personalization}. Some observations can be obtained as follows: 

\begin{itemize}[leftmargin=*]
    \item \textbf{Jaccard Similarity (0.7605):} The high Jaccard similarity coefficient indicates that the recommendation results before and after applying RETURN are highly consistent for benign users. This suggests that RETURN preserves the majority of the original recommendations, although some items are removed or replaced. 
    \item \textbf{Common Items Ratio (0.8706):} The proportion of items that co-occur in both the benign and RETURN-processed recommendations is 87.06\%. This further demonstrates that RETURN maintains the core set of recommended items, ensuring minimal disruption to the personalized recommendations.
    \item \textbf{Shannon Entropy:} The Shannon entropy values for both the benign (9.8616) and RETURN-processed (9.8292) recommendations are nearly identical. This indicates that RETURN does not significantly reduce the diversity of the recommendations, preserving the richness and variety of the suggested items. 
\end{itemize}

\begin{table}[htbp]
  \centering
  \caption{Impact of RETURN on the personalization of recommendations}
    \scalebox{0.9}{
    \begin{tabular}{c|c|c|c}
    \toprule
          & Shannon Entropy & Jaccard Similarity & Common Items Ratio \\
    \hline
    Benign & 9.8616 & \multirow{2}[0]{*}{0.7605} & \multirow{2}[0]{*}{0.8706} \\
    RETURN & 9.8292 &       &  \\
    \bottomrule
    \end{tabular}%
    }
  \label{tab:impact_to_personalization}%
\end{table}%

%% file: Section/Conclusion.tex
\section{Conclusion}\label{sec:conclusion}
In this paper, we propose a novel framework RETURN by retrieving collaborative knowledge from external databases to enhance the robustness of existing LLM-empowered RecSys in a plug-and-play manner. 
Specifically, the proposed RETURN first converts the user interactions within external databases into collaborative item graphs to implicitly encode the collaborative signals. 
Then, the potential perturbations are located by retrieving relevant knowledge from the generated graphs. 
To mitigate the negative impact of perturbations and maintain the integrity of user preference, a retrieval-augmented denoising strategy is introduced to purify the input user profile. 
Finally, a robust ensemble recommendation method is proposed to generate the final recommendations by adopting a decision fusion strategy. 
Comprehensive experiments on real-world datasets demonstrate the effectiveness of the proposed RETURN and highlight the potential of introducing external collaborative knowledge to enhance the robustness of LLM-empowered RecSys.